\documentclass[submission, Phys]{SciPost}

\hypersetup{
    colorlinks,
    linkcolor={red!50!black},
    citecolor={blue!50!black},
    urlcolor={blue!80!black}
}

\usepackage{array}
\usepackage{multirow}

\usepackage[bitstream-charter]{mathdesign}
\DeclareSymbolFont{usualmathcal}{OMS}{cmsy}{m}{n}
\DeclareSymbolFontAlphabet{\mathcal}{usualmathcal}

\fancypagestyle{SPstyle}{
\fancyhf{}
\lhead{\colorbox{scipostblue}{\bf \color{white} ~SciPost Physics }}
\rhead{{\bf \color{scipostdeepblue} ~Submission }}

\fancyfoot[C]{\textbf{\thepage}}
}

\begin{document}

\pagestyle{SPstyle}

\begin{center}{\Large \textbf{\color{scipostdeepblue}{
%%%%%%%%%% TODO: Write your article's title here
Boundary Conditions for Extremal Black Holes from 2d Gravity
}}}\end{center}

\begin{center}\textbf{
% Use (full) first name (+ middle name initials) + surname format.
% Separate subsequent authors by a comma, omit comma and use "and" for the last author.
% Mark the corresponding author(s) with a superscript symbol in this order
% \star, \dagger, \ddagger, \circ, \S, \P, \parallel, ...
St\'ephane Detournay\textsuperscript{1$\star$},
Thomas Smoes\textsuperscript{1$\dagger$} and
Raphaela Wutte\textsuperscript{1,2$\ddag$}
}\end{center}

\begin{center}
{\bf 1} Physique Mathématique des Interactions Fondamentales, Université Libre de Bruxelles, Campus Plaine - CP 231, 1050 Bruxelles, Belgium
\\
{\bf 2} Department of Physics and Beyond: Center for Fundamental Concepts in Science, Arizona State
University, Tempe, Arizona 85287, USA
%%%%%%%%%% END TODO: AFFILIATIONS
% Provide email address of corresponding author(s)
\\[\baselineskip]
$\star$ \href{mailto:email1}{\small sdetourn@ulb.ac.be}\,,\quad
$\dagger$ \href{mailto:email2}{\small thomas.smoes@ulb.be} \,,\quad
$\ddag$ \href{mailto:email2}{\small rwutte@hep.itp.tuwien.ac.at}
\end{center}

\section*{\color{scipostdeepblue}{Abstract}}
\textbf{%
\textbf{We devise new boundary conditions for the near-horizon geometries of extremal BTZ and Kerr black holes, as well as for the ultra-cold limit of the Kerr-de Sitter black hole. These boundary conditions are obtained as the higher-dimensional uplift of recently proposed boundary conditions in two-dimensional gravity. Their asymptotic symmetries consist in the semi-direct product of a Virasoro and a current algebra, of which we determine the central extensions.}
}

\vspace{\baselineskip}

%%%%%%%%%% BLOCK: Copyright information
% This block will be filled during the proof stage, and finilized just before publication.
% It exists here only as a placeholder, and should not be modified by authors.
% \noindent\textcolor{white!90!black}{%
% \fbox{\parbox{0.975\linewidth}{%
% \textcolor{white!40!black}{\begin{tabular}{lr}%
%   \begin{minipage}{0.6\textwidth}%
%     {\small Copyright attribution to authors. \newline
%     This work is a submission to SciPost Physics. \newline
%     License information to appear upon publication. \newline
%     Publication information to appear upon publication.}
%   \end{minipage} & \begin{minipage}{0.4\textwidth}
%     {\small Received Date \newline Accepted Date \newline Published Date}%
%   \end{minipage}
% \end{tabular}}
% }}
% }
%%%%%%%%%% BLOCK: Copyright information

% For convenience during refereeing we turn on line numbers:
%\linenumbers

\vspace{10pt}
\noindent\rule{\textwidth}{1pt}
\tableofcontents
\noindent\rule{\textwidth}{1pt}
\vspace{10pt}

% Write your article contents here, starting from first \section.
% An example structure is given below.

\section{Introduction and Outlook}

When formulating a physical problem, the equations of motion have to be supplemented by boundary conditions (BCs) on the dynamical variables. In fact, the latter turn out to be as important as the former \cite{Fok1959-FOKTTO} (cited in \cite{Bunster:2014mua}). 
This is especially clear when the theory is formulated in terms of an action principle and the partition function defined through a path integral: the boundary conditions specify the off-shell configurations over which the integral has to be performed. Systems with identical field equations but different boundary conditions could describe significantly distinct physical phenomena and exhibit different contents (e.g. closed/open strings, Dirichlet vs Neumann BCs).

Boundary conditions play a crucial role in gauge theories, in particular in theories of gravity. There, the set of metrics satisfying given equations of motion and boundary conditions constitute the configuration space of the theory, which can be identified with its phase space. 
The identification of the symmetries of the phase space are of crucial importance since one expects, upon quantization, that the Hilbert space of the corresponding quantum theory will fall into a representation of the symmetry group, for instance in the spirit of the geometric quantization program \cite{Kostant1970QuantizationAU, MR0260238}.

 In gauge theories, the symmetries of the phase space, mapping one solution onto another with distinct physical charges, are of great importance. These are called {\itshape asymptotic symmetries} and form the {\itshape asymptotic symmetry group} (ASG). The study of asymptotic symmetries in gravity theories has a long history that started in 1962 with the founding papers
 \cite{Bondi:1962px, Sachs:1962wk}
 which identified the BMS group of supertranslations and Lorentz transformations as ASG of four-dimensional asymptotically flat spacetimes. It was later extended to include superrotations in \cite{deBoer:2003vf, Barnich:2009se,Barnich:2010eb}
 and diffeomorphisms on the 2-sphere in \cite{Campiglia:2014yka, Campiglia:2015yka}.
 The renewed interest in BMS symmetries is largely due to 
recent work on BMS invariance of scattering amplitudes \cite{Strominger:2013jfa}
and the ``infrared triangle" relating BMS supertranslation symmetries, Weinberg’s soft graviton theorem and the displacement memory effect
\cite{Strominger:2017zoo}.

Equally impactful is the discovery by Brown and Henneaux of two-dimensional conformal symmetry in the asymptotic structure of AdS$_3$ gravity\cite{Brown:1986nw}, an early precursor of the AdS/CFT correspondence \cite{Maldacena:1998bw}. It brought deep insights into the holographic nature of gravity and in particular the identification of microscopic degrees of freedom for specific classes of black holes, either asymptotically AdS$_3$ (the BTZ black hole \cite{Banados:1992gq, Banados:1992wn}) \cite{Strominger:1998yg} or with an AdS$_3$ factor in their near-horizon geometry \cite{Cvetic:1998xh}.
The three-dimensional situation in flat space has been addressed more recently, identifying the BMS$_3$ asymptotic symmetry algebra at null infinity 
\cite{Ashtekar:1996cd, Barnich:2006av} 
and at spatial infinity 
\cite{Compere:2017knf}.
The flat limit from AdS$_3$ to Minkowski was described in 
\cite{Barnich:2006av}
for the symmetry algebra, and for the full phase space in 
\cite{Barnich:2012aw}. 
The flat spacetime cosmologies 
\cite{Cornalba:2002fi, Cornalba:2003kd}
-- the flat counterparts of the BTZ black holes -- and their thermodynamical interpretation in terms of BMS$_3$ symmetries were addressed in \cite{Barnich:2012xq, Bagchi:2012xr}.
Interestingly, the non-uniqueness of the ASG given a vacuum solution 
and non-trivial zero-mode solutions
has been brought to light only rather recently. Superrotations in four-dimensional asymptotically flat space have been introduced almost half a century after the works of Bondi, van der Burg, Metzner and Sachs. 
In three-dimensional gravity, a variety of alternative boundary conditions -- allowing e.g. for a fluctuating boundary metric, in contrast with the Dirichlet-like Brown-Henneaux boundary conditions -- have been proposed in recent years both for AdS$_3$ 
\cite{Porfyriadis:2010vg, Compere:2013aya, Afshar:2016wfy, Troessaert:2013fma, Avery:2013dja, Grumiller:2016pqb}
and Minkowski space 
\cite{Afshar:2016kjj, Detournay:2016sfv, Grumiller:2017sjh}
exhibiting in general different ASGs, hence potentially different field theory dual interpretations.  A particular way of relating different ASGs in three dimensions has been discussed in \cite{Grumiller:2019ygj}.

Among holographic dualities involving AdS spaces, the two-dimensional case has always stood out as more challenging. The boundary of AdS$_2$ consists in two disconnected pieces, and finite energy excitations have been observed to destroy the asymptotic geometry \cite{Fiola:1994ir,Maldacena:1998uz}. This has long been a hindrance for a microscopic understanding of extremal higher-dimensional black holes, as these generally exhibit a near-horizon geometry including an AdS$_2$ factor \cite{Kunduri:2013gce, Dunajski:2023xrd} when the cosmological constant is non-positive (we will later discuss a situation where AdS$_2$ is replaced by Mink$_2$ for the near-horizon limit of the ultra-cold Kerr-de Sitter black hole \cite{Anninos:2009yc}).  
It has however recently been found how to circumvent these obstructions and identify the relevant degrees of freedom describing the low energy physics driving a black hole away from extremality. It consists in considering {\itshape nearly}-AdS$_2$ holography by including the leading corrections away from pure AdS$_2$ \cite{Almheiri:2014cka, Maldacena:2016upp} (for reviews, see e.g. 
\cite{Sarosi:2017ykf, Mertens:2022irh} 
or App.B of \cite{Castro:2021csm}).
The physics near the horizon of near-extremal black holes in higher dimensions can be shown to be universally described by a particular occurrence of two-dimensional dilaton gravity theory --  JT gravity \cite{Teitelboim:1983ux, Jackiw:1984je}, with certain Dirichlet boundary conditions at the boundary of AdS$_2$. The latter exhibit time-reparametrization invariance whose generators\footnote{Note that the reparameterisation symmetry is broken both spontaneously by pure AdS$_2$ and explicitly due to the non-trivial boundary condition for the dilaton} are reminiscent of (one half of) the Brown-Henneaux ones \cite{Hotta:1998iq,Cadoni:1999ja,Navarro-Salas:1999zer}.
Again, like in higher dimensions, different sets of boundary conditions with different symmetries can be considered \cite{Grumiller:2017qao}. 
Recently, new boundary conditions for AdS$_2$ have been proposed \cite{Godet:2020xpk}, where the usual time-reparametrization symmetry is enhanced  with an additional local U(1) symmetry, extending the symmetry algebra to a Virasoro-Kac-Moody $U(1)$ algebra.
The latter represent the symmetries of a so-called Warped CFT (WCFT) \cite{Hofman:2011zj, Detournay:2012pc}, a two-dimensional non-relativistic theory with chiral scale invariance and $SL(2,R)\times U(1)$ global symmetry (see \cite{Compere:2013aya, Castro:2015uaa, Castro:2015csg, Song:2017czq, Apolo:2018eky, Chaturvedi:2018uov, Aggarwal:2022xfd} for some of their properties). 

The goal of the present work will be to explore new boundary conditions for extremal black holes, in particular determine whether the boundary conditions of \cite{Godet:2020xpk} and \cite{Afshar:2019axx} can be uplifted to the near-horizon geometry in higher dimensions. Our work can thus be regarded as a proof of principle that certain boundary conditions existing in 2d gravity have a natural uplift to higher dimensions.

Motivations stem from the ubiquity of AdS$_2$ in the near-horizon geometry of extremal black holes, but also from the Kerr/CFT correspondence\cite{Guica:2008mu} -- an attempt to relate four-dimensional extremal Kerr black holes to a chiral CFT in two dimensions. The argument there parallels the connection between AdS$_3$ and 2d CFTs, where the AdS$_3$ near-region throat geometry is replaced with the NHEK (near-horizon
extreme Kerr) geometry found by Bardeen and Horowitz \cite{Bardeen:1999px} via a near-horizon limit. Constant polar sections of the NHEK geometry consist in deformations of AdS$_3$, termed Warped AdS$_3$ (WAdS$_3$) spaces \cite{Moussa:2003fc, Bouchareb:2007yx, Banados:2005da, Detournay:2012dz, Tonni:2010gb, Donnay:2015iia, Detournay:2016gao},
where the original undeformed $SO(2,2)$ isometries get broken down to $SL(2,R)\times U(1)$. Holographic properties of WAdS$_3$ spaces have been explored over the years
\cite{Detournay:2012pc, Compere:2007in, Compere:2008cv, Compere:2009zj, Henneaux:2011hv, Song:2016gtd, Azeyanagi:2018har, Aggarwal:2020igb, Anninos:2008fx, Chen:2009cg, Song:2016pwx, Compere:2014bia, Guica:2013jza, El-Showk:2011euy, Song:2011sr, Guica:2017lia, Bzowski:2018pcy, Aggarwal:2023peg}
as a toy model for Kerr black holes. For generic Kerr black holes, the relevance of WCFTs was pointed out in \cite{Aggarwal:2019iay} in the spirit of \cite{Haco:2018ske}. In the extremal limit, the question is still open.

The Kerr/CFT proposal is based on boundary conditions extending the $U(1)$ part of the isometry group into a Virasoro algebra, whose computed central charge allowed to reproduce the macroscopic Bekenstein-Hawking extremal Kerr entropy. This was one of the landmarks of the original proposal\footnote{This is currently being debated in recent works suggesting instead a vanishing entropy at low temperatures \cite{Rakic:2023vhv, Kapec:2023ruw}.}. From a gravity perspective these boundary conditions might seem unnatural, as their symmetries do not include all the exact symmetries of the background. Soon after the Kerr/CFT proposal, other boundary conditions have been proposed extending instead the $SL(2,R)$ part of the isometries, but found vanishing central extensions \cite{Matsuo:2009sj, Matsuo:2009pg}. In this work, we will propose new boundary conditions for the NHEK geometry, 
inspired by the Godet-Marteau analysis in two dimensions \cite{Godet:2020xpk}.
One feature of these boundary conditions and their symmetries is the dependence of the generators on (retarded) time. Extracting a non-trivial symmetry algebra therefore requires to integrate charges over time instead of the usual constant-time, angular integration. This procedure has been applied both in two and higher dimensions \cite{Cadoni:1998sg, Cadoni:1999ja, Grumiller:2017qao, Afshar:2015wjm}. Integration over time produces time-averaged charges which can be seen to give a canonical representation of the asymptotic symmetry algebra with non-trivial central extensions. 
The procedure can also be interpreted from the boundary perspective, in particular when the putative dual theory is two-dimensional (CFT, WCFT, or other) and enjoys modular invariance. A modular invariant field theory at finite chemical potentials is naturally defined on a torus with two cycles, the spatial one (angular identifications) and the thermal one (in particular, time has a 
period set by the inverse temperature). Its partition function can be expressed either as a trace over states defined on spatial cycles (with charges integrated over a spatial cycle) and evolved with the usual hamiltonian operator, or as states defined on thermal cycles (hence with time periodic in particular and charges integrated over a thermal cycle) and evolved with the angular momentum operator. This yields
one possible boundary interpretation of a bulk time integration and it is this interpretation that we will employ throughout this work.

The paper is organized as follows. As a warm-up, we devise in Sect. 2 new boundary conditions for the near-horizon limit of extremal BTZ black holes, the so-called selfdual orbifold. Kerr/CFT-like boundary conditions had appeared e.g. in \cite{Balasubramanian:2009bg}. Here we define a new phase space with WCFT symmetries of which we identify the non-trivial central extensions, the Virasoro one coinciding with the Brown-Henneaux central charge. In Sect. 3 we turn to boundary conditions including the NHEK geometry. Following a similar strategy, we define a phase space, identify their asymptotic symmetries, and compute the asymptotic charges. The latter are shown to satisfy through their Poisson bracket a WCFT algebra with non trivial central extensions both for the Virasoro and current algebra. The Virasoro central extensions is seen to match that of the original Kerr/CFT correspondence. We address a slightly different case in Sect. 4. It consists in boundary conditions including the near-horizon limit of the ultra-cold Kerr-de Sitter black hole in 4 dimensions (where the 3 horizons come to coincide). There is no known way to associate a CFT or any other boundary theory for that matter to the ultracold limit \cite{Anninos:2009yc} (see however \cite{Castro:2022cuo} studying the response of ultracold black holes to small perturbations). The latter does not fall in the general category of AdS$_2$ near-horizon geometry. Instead, the AdS$_2$ factor is replaced by two-dimensional Minkowski space. As it turns out, boundary conditions for Mink$_2$ have been proposed and their asymptotic symmetries determined \cite{Afshar:2019axx, Afshar:2021qvi}. We uplift these boundary conditions to 4 dimensions, demonstrating that they yield well defined charges and asymptotic symmetry algebras, again consisting in a WCFT algebra of which we compute the central extensions. This provides a first step towards building a holographic dual for ultracold Kerr-dS black holes.
In section \ref{sec:conclusion} we conclude with a summary and interpretation of our results in the context of holography.

\section{Extremal BTZ}\label{sec:BTZ}

\subsection{Geometry and Near-horizon Limit}\label{}
The metric of the extremal BTZ black hole is
\begin{equation}
    ds^2 = - \frac{(r^2 -r^2 _h )^2}{r^2} dt^2 + \frac{r^2}{(r^2 - r^2 _h )^2} dr^2 + r^2 \left( d\phi - \frac{r^2 _h}{r^2} dt \right)^2
\,,
\end{equation}
where $r_h$ is the horizon radius and where the AdS radius $l$ has been set to one. We consider the change of coordinates
\begin{equation}
    t = \frac{\tau}{\epsilon} \,, \quad r^2 = r^2 _h + \epsilon \rho \,, \quad \phi = \varphi + \frac{\tau}{\epsilon}
\end{equation}
and then study the near-horizon limit (NHL) 
by taking $\epsilon \rightarrow 0$. The extremal BTZ metric becomes \cite{deBoer:2010ac}
\begin{align}
\label{beforeresc}
    ds^2 & = \frac14 \frac{d\rho^2}{\rho ^2} + 2\rho d\tau d\varphi + r^2 _h d\varphi ^2 \nonumber \\
    & = \frac14 \frac{d\rho^2}{\rho ^2} - \frac{\rho ^2}{r^2 _h} d\tau ^2 + r^2 _h \left( d\varphi + \frac{\rho}{r^2 _h} d\tau \right)^2.
\end{align}
In order to apply Godet-Marteau boundary conditions on this metric, we will write it in a system of coordinates that is similar to the Bondi gauge described in \cite{Godet:2020xpk} for AdS$_2$. We thus define new coordinates $(u, \hat{r}, \hat{\varphi})$ such that
\begin{equation}
\label{rescalingbf}
    \tau = \frac{u}{2} - \frac{1}{2 \hat{r}} \,, \quad \rho = r_h \hat{r} \,, \quad \varphi = \frac{1}{2 r_h} (\hat{\varphi} - \ln \hat{r})
\end{equation}
and the metric becomes
\begin{equation}
    ds^2 = \frac14 (-\hat{r}^2 du^2 - 2 du d\hat{r}) + \frac14 (\hat{r} du + d\hat{\varphi})^2 . \label{btz_bondi 2.1}
\end{equation}
From now on, we will omit "\textasciicircum " of the coordinates, keeping in mind that the new coordinates are different from the ones in \eqref{beforeresc}.

\subsection{Phase Space and Asymptotic Killing Vectors}\label{}
Inspired by the Godet-Marteau boundary conditions for AdS$_2$ \cite{Godet:2020xpk}, we consider the following family of metrics%(\ref{btz_bondi 2.1}) 
\begin{subequations}
    \label{bc_btz 2.2}
\begin{align}
    ds^2 &= \frac14 \bigg( (-r^2 + 2P(u)r + 2T(u) ) du^2 - 2 du dr \bigg) + \frac14 (r du + d\varphi )^2\,,   \\
    &= ds^2_{2d} + \frac14 (r du + d\varphi )^2
\end{align}
\end{subequations}
where $P$ and $T$ are arbitrary functions of $u$. 
Here, the first part of the metric $ds^2_{2d}$ corresponds to boundary conditions that were previously imposed for 2d gravity \cite{Godet:2020xpk}. 
%depend on the functions $\mathcal{F}$ and $\mathcal{G}$.
The boundary conditions \eqref{bc_btz 2.2}
can be obtained from (\ref{btz_bondi 2.1}) by applying the finite coordinate transformation
\begin{equation}
    u \rightarrow \mathcal{F} (u) \,, \quad r \rightarrow \frac{1}{\mathcal{F}'} (r+\mathcal{G}' (u) ) \,, \quad \varphi \rightarrow \varphi - \mathcal{G} (u) \,. \label{finite_transf 2.2}
\end{equation}
The functions $P, T, \mathcal{F}$ and $\mathcal{G}$ are related by
\begin{align}
    P(u) & = -\mathcal{G}' (u) + \frac{\mathcal{F}''(u)}{\mathcal{F}'(u)} \,, \\
    T(u) & = -\frac12 \mathcal{G}' (u)^2 + \mathcal{G}' (u) \frac{\mathcal{F}''(u)}{\mathcal{F}'(u)} - \mathcal{G}''(u) \,.
\end{align}
The asymptotic Killing vectors generating the transformations (\ref{finite_transf 2.2}) are given by
\begin{equation}
    \xi = \epsilon (u) \partial _u + (-r\epsilon ' (u) - \zeta ' (u)) \partial _r + \zeta (u) \partial _\varphi \,, \label{akv 2.2}
\end{equation}
where $\epsilon (u)$ and $\zeta (u)$ are two arbitrary functions of $u$. By applying the Lie derivative on the metric (\ref{bc_btz 2.2}), we can also find the variations of $P(u)$ and $T(u)$
\begin{align}
    \delta _\xi P & = \epsilon P' + \epsilon ' P + \epsilon '' + \zeta ' \,, \\
    \delta _\xi T & = \epsilon T' + 2 \epsilon ' T - \zeta ' P + \zeta '' \,. 
\end{align}
Alternatively, we can define a perturbation $h_{\mu \nu}$ on the background metric (\ref{btz_bondi 2.1}) such that
\begin{subequations}
\label{falloff}
\begin{align}
    & h_{uu} = \mathcal{O} (r) \,, \quad h_{ur} = \mathcal{O} (r^{-2}) \,, \quad h_{u \varphi} = \mathcal{O} (r^{-1}) \,, \\
    & h_{rr} = \mathcal{O} (r^{-3}) \,, \quad h_{r \varphi} = \mathcal{O} (r^{-2}) \,, \quad h_{\varphi \varphi} = \mathcal{O} (r^{-1})
\end{align}
\end{subequations}
and the vectors solving the asymptotic Killing equation are given by
\begin{equation}
\label{asymptKillingBTZ}
    \xi = (\epsilon (u) + \mathcal{O} (r^{-2})) \; \partial _u + (-r\epsilon ' (u) - \zeta ' (u) + \mathcal{O} (r^{-1})) \; \partial _r + (\zeta (u) + \mathcal{O} (r^{-2})) \; \partial _\varphi\, .
\end{equation}
Fixing the coordinate system, by setting $g_{u r} = -1/4$, $g_{r r}=0$ and $g_{r \varphi} = 0$, and assuming that the remaining components admit an expansion in powers of $r$
\begin{subequations}
\begin{align}
g_{uu} &= r g_{uu1}(u, \varphi) +g_{uu0}(u, \varphi) + \mathcal{O}(r^{-1})\,, \\
g_{u \varphi} &= \frac{r}{4} + \frac{g_{u \varphi 0}(u, \varphi)}{r} + \mathcal{O}(r^{-2})\,, \\
g_{\varphi \varphi} &= \frac{1}{4} + \frac{g_{\varphi \varphi {-1}}(u, \varphi)}{r}+ \mathcal{O}(r^{-2})\,,
\end{align}
\end{subequations}
one readily obtains that 
\eqref{bc_btz 2.2} is the unique class of metrics that solves the vacuum Einstein equations with a negative cosmological constant and  the fall-off conditions \eqref{falloff}. It is in this sense, that \eqref{falloff} and \eqref{asymptKillingBTZ} are equivalent to \eqref{bc_btz 2.2} 
and \eqref{akv 2.2}.
In the following, we will always work with a class of metrics instead of directly working with boundary conditions.

From now on, we assume that $u$ is periodic with period $L \in i \mathbb{R}$, where $i$ is the imaginary unit, and define the modes of the vectors \eqref{akv 2.2} as 
\begin{equation}
    l_n = \xi \left(\epsilon = \frac{L }{2\pi} e^{2\pi i n u /L} , \; \zeta = 0\right) , \quad j_n = \xi \left(\epsilon = 0, \; \zeta = \frac{L}{2\pi i} e^{2\pi i n u /L} \right),
\end{equation}
where $n \in \mathbb{Z}$. 
The motivation for this arises from the Euclidean where Euclidean time is periodic with period $\beta$. Wick rotating, the period of Lorentzian time becomes $L = i \beta$ where $\beta$ is the temperature.
These modes satisfy a warped Witt algebra under the Lie bracket:
\begin{subequations}
\label{warpedWitt}
\begin{align}
    & i [l_m , l_n ] = (m-n) l_{m+n}\,, \label{lie_ll 2.2}\\
    & i [l_m , j_n] = -n j_{m+n}\,, \label{lie_lj 2.2} \\
    & i [j_m , j_n] = 0\,. \label{lie_jj 2.2}
\end{align}
\end{subequations}

\subsection{Charge Algebra}\label{}
The infinitesimal charge difference between two geometries $g_{\mu \nu}$ and $g_{\mu \nu} + h_{\mu \nu}$, where $h_{\mu \nu}$ is an infinitesimal perturbation, is given by
\begin{equation}
    \delta Q_\xi [h , g] = \int _{\partial \Sigma} k_\xi [h , g]\,. \label{charge_diff 2.3}
\end{equation}
The differential form $k_\xi$ associated to an asymptotic Killing vector $\xi$ is defined by\footnote{See e.g. \cite{Compere:2019qed} for a pedagogical account and references.} 
\begin{align}
    k_\xi [h , g] = & \; \frac{\sqrt{-g}}{8\pi G} (d^{n-2} x)_{\mu \nu} \Bigl( \xi ^\mu \nabla _\sigma h^{\nu \sigma} - \xi ^\mu \nabla ^\nu h + \xi _\sigma \nabla ^\nu h^{\mu \sigma} 
     + \frac12 h \nabla ^\nu \xi ^\mu - h^{\rho \nu} \nabla _\rho \xi ^\mu \Bigr)\,,
\end{align}
where $n$ is the space-time dimension, $\nabla$ is the covariant derivative of $g_{\mu \nu}$ and $ h = g^{\mu \nu} h _{\mu \nu}$. 
One readily checks that integrating (\ref{charge_diff 2.3}) along the direction of $\varphi$ 
over a constant $u$ surface and taking the limit $r \rightarrow \infty$, yields zero -- all surface charges vanish.
One may obtain non-zero surface charges by integrating (\ref{charge_diff 2.3}) along the direction of $u$ over 
a constant $\varphi$ surface and then taking the limit $r \rightarrow \infty$, which is what we will do in what follows.
The motivation for this stems from holography. In particular, as detailed in the introduction, assuming that the putative dual field theory is modular invariant, we can use this modular invariance to switch the angular and temporal cycle for the computation of the charges.
\\  \indent
For this, we begin by defining the variation of the metric (\ref{bc_btz 2.2}) as
\begin{equation}
    h_{\mu \nu} \equiv \delta g_{\mu \nu} = \frac{\partial g_{\mu \nu}}{\partial P} \delta P + \frac{\partial g_{\mu \nu}}{\partial T} \delta T\,.
\end{equation}
Computing the variation of the charges, we find 
\begin{equation}
    \delta Q_\xi = \frac{1}{16 \pi G} \int ^{L} _0 du (\epsilon  \delta T - \zeta \delta P)\,.
\end{equation}
We see that this expression can be directly integrated in order to obtain the finite charges
\begin{equation}
    Q_\xi = \frac{1}{16 \pi G} \int ^{L} _0 du (T(u) \epsilon (u) - P(u) \zeta (u))\,,
\end{equation}
where the metric of the extremal black hole in the NHL, which has $P(u) = T(u) = 0$, has been chosen as the background metric. In particular, we define
\begin{align}
    L_n & = Q_{l_n} = \frac{1}{16 \pi G} \int ^{L} _0 du \,T(u) \frac{L}{2\pi} e^{2\pi i n u /L },  \\
    J_n & = Q_{j_n} = -\frac{1}{16 \pi G} \int ^{L} _0 du\, P(u) \frac{L }{2\pi i} e^{2\pi i n u /L},
\end{align}
Computing the algebra of these charges under the Dirac bracket yields
\begin{subequations}
\label{WCFTalgebrabefore}
\begin{align}
    i \{ L_m , L_n \} & = i\delta _{l_n} L_m = (m-n) L_{m+n} \,, \label{dirac_LL 2.3} \\
    i \{ L_m , J_n \} & = i\delta _{j_n} L_m  = - n J_{m+n} - \frac{L}{16\pi G} m^2 \delta _{m+n,0}\,, \label{dirac_LJ 2.3} \\
    i \{ J_m , J_n \} & = i\delta _{j_n} J_m = \frac{ L^2}{32 \pi ^2 G} m \delta _{m+n, 0}\,. \label{dirac_JJ 2.3}
\end{align}
\end{subequations}

The algebra described by the relations \eqref{WCFTalgebrabefore} corresponds to a Virasoro-Kac-Moody $U(1)$ algebra, the symmetry algebra of a WCFT,
\begin{subequations}
\label{WCFTalgebra}
\begin{align}
    i \{L_m , L_n \} & = (m-n) L_{m+n} + \frac{c}{12} m^3 \delta _{m+n,0}\,, \label{WCFT_LL 2.3} \\
     i \{L_m , J_n \} & = -n J_{m+n} - i \kappa m^2 \delta _{m+n,0}\,, \label{WCFT_LJ 2.3}\\
    i \{J_m , J_n \} & = \frac{k}{2} m \delta _{m+n,0}\,. \label{WCFT_JJ 2.3}
\end{align}
\end{subequations}
with central charges $c$, $\kappa$ and $k$  
\begin{equation}
    c=0\,, \quad \kappa = \frac{L}{16 \pi i G}\,, \quad k = \frac{L^2}{16 \pi ^2 G}\,. \label{central_charges 2.3}
\end{equation}
Note that the central extensions obtained here are manifestly real, since $L \in i \mathbb{R}$.

\subsection{Boundary Conditions in Schwarzschild-like Coordinates}\label{BC_Schwarzschild_BTZ}
Previously, we applied the Godet-Marteau boundary conditions on the extremal BTZ black hole by introducing a new system of coordinates (with a retarded time $u$). With this system of coordinates, the metric was written in a form that was similar to the Bondi gauge for AdS$_2$. Here, we perform our analysis in the Schwarzschild-like system of coordinates. In these coordinates, the metric of the extremal BTZ black hole (in the NHL) reads \eqref{beforeresc}.
Upon rescaling $\rho \rightarrow (r_h \rho) /2$ and $\varphi \rightarrow \varphi / (2 r_h)$, the metric \eqref{beforeresc} becomes
\begin{equation}
    ds^2 = \frac14 \left( \frac{d\rho ^2}{\rho^2} - \rho^2 d\tau^2 \right) + \frac14 \left( d\varphi + \rho d\tau \right)^2 .
\end{equation}
We now impose Godet-Marteau boundary conditions on this metric by applying a finite coordinate transformation given by 
\begin{equation}
    \tau \rightarrow \mathcal{F} (\tau) \,, \quad \rho \rightarrow \frac{1}{\mathcal{F}'} (\rho + \mathcal{G}' (\tau )) \,, \quad \varphi \rightarrow \varphi - \mathcal{G} (\tau) \,. \label{finite_transf 2.4}
\end{equation}
Defining a function $\mathcal{H} (\tau) \equiv \mathcal{F}''(\tau) / \mathcal{F}'(\tau)$, this transformation yields the metric components 
\begin{subequations}
\label{bccomponents2.4}
\begin{align}
    g_{\tau \tau} &=  - \frac12 \rho \mathcal{G}' (\tau) - \frac14 \mathcal{G}' (\tau)^2 + \frac{((\rho + \mathcal{G}'(\tau) ) \mathcal{H} (\tau) - \mathcal{G} '' (\tau) )^2}{4 (\rho + \mathcal{G}' (\tau))^2} \,, \label{g_tau_tau 2.4}\\
    g_{\tau \rho} &=  \frac{-(\rho + \mathcal{G}'(\tau) ) \mathcal{H} (\tau) + \mathcal{G}'' (\tau)}{4 (\rho + \mathcal{G}'(\tau))^2} \,, \quad g_{\tau \varphi} =  \frac{\rho}{4} \,, \\
    g_{\rho \rho } &=  \frac{1}{4 (\rho + \mathcal{G}'(\tau))^2} \,, \quad g_{\rho \varphi} = 0 \,, \quad g_{\varphi \varphi} = \frac14 \,. \label{g_rho_rho 2.4}
\end{align}
\end{subequations}
The metric of the extremal BTZ black hole (in the NHL) corresponds to \eqref{bccomponents2.4} with $\mathcal{G}' (\tau ) = 0$ and $\mathcal{H} (\tau) = 0$.

The asymptotic Killing vectors generating the transformations (\ref{finite_transf 2.4}) are given by
\begin{equation}
    \xi = \epsilon (\tau ) \partial _\tau - (\rho \epsilon ' (\tau) + \zeta ' (\tau )) \partial _\rho  + \zeta (\tau) \partial _\varphi \,, \label{akv 2.4}
\end{equation}
where $\epsilon (\tau )$ and $\zeta (\tau)$ are two arbitrary functions of $\tau$. By applying the Lie derivative on the metric, we find the variations of $\mathcal{G}' (\tau )$ and $\mathcal{H} (\tau)$:
\begin{align}
    & \delta _{\xi} \mathcal{G}' (\tau) = \epsilon '(\tau) \mathcal{G}' (\tau) + \epsilon (\tau) \mathcal{G}'' (\tau) - \zeta '(\tau) \,, \\
    & \delta _{\xi} \mathcal{H} (\tau) = \epsilon '(\tau) \mathcal{H} (\tau) + \epsilon '' (\tau) + \epsilon (\tau) \mathcal{H} '(\tau) \,.
\end{align}
In the following, we assume that $\tau$ is periodic with period $L$. We define modes as
\begin{equation}
    l_n = \xi \left(\epsilon = \frac{L }{2\pi} e^{2\pi i n \tau /L} , \; \zeta = 0\right) , \quad j_n = \xi \left( \epsilon = 0, \; \zeta = \frac{L}{2\pi i} e^{2\pi i n \tau /L} \right).
\end{equation}
Under the Lie bracket, these modes satisfy the warped Witt algebra \eqref{warpedWitt}.

Here, the integral considered in (\ref{charge_diff 2.3}) for the computation of the charges is taken over $\tau$ while $\rho \rightarrow \infty$ and $\varphi$ is constant. Explicitly computing the charges yields
\begin{equation}
    Q_\xi = \frac{1}{32 \pi G} \int _0 ^L d\tau \left( 2 \zeta (\tau) \mathcal{G}' (\tau ) - \epsilon (\tau) \mathcal{G}' (\tau )^2 + 2 \epsilon '(\tau) \mathcal{H} (\tau) + \epsilon (\tau) \mathcal{H} (\tau)^2 \right),
\end{equation}
where the metric of the extremal black hole in the NHL, which has $\mathcal{G}' (\tau) = \mathcal{H} (\tau) = 0$, has been chosen as the background metric. We define
\begin{align}
    & L_n = Q_{l_n} = \frac{1}{32 \pi G } \int _0 ^{L} d\tau \left( - \mathcal{G}' (\tau) ^2 + \frac{4\pi i n }{L} \mathcal{H} (\tau) + \mathcal{H} (\tau)^2 \right) \frac{L }{2\pi} e^{2\pi i n \tau /L}\,, \\
    & J_n = Q_{j_n} = \frac{1}{32 \pi G } \int _0 ^{L} d\tau ( 2 \mathcal{G}' (\tau) ) \frac{L}{2\pi i} e^{2\pi i n \tau /L}\,.
\end{align}
The algebra of the charges $L_n$ and $J_n$ is given by (\ref{WCFT_LL 2.3})-(\ref{WCFT_JJ 2.3}) with central charges 
\begin{equation}
    c^* = \frac{3}{2G} \,, \quad  \kappa ^* = 0 \,, \quad k^* = \frac{L^2}{16 \pi ^2 G} \,, \label{central_charges 2.4}
\end{equation}
which are different from those found in (\ref{central_charges 2.3}).
We would like to know if it is possible to relate % \sout{ the central charges}
algebras  \eqref{WCFTalgebra} with different central charges, given by
(\ref{central_charges 2.3}) and (\ref{central_charges 2.4}), respectively. By defining new surface charges \cite{Godet:2020xpk}
\begin{equation}
\label{changeinalgebra}
    L_n^* := L_n + \frac{2 i \kappa}{k} n J_n
\end{equation}
it is possible to go from an algebra with central charges $c$, $\kappa$ and $k$ to a new algebra with central charges given by
\begin{equation}
    c^* = c - \frac{24 \kappa ^2}{k} \,, \quad \kappa ^* = 0 \,, \quad k^* = k \,. \label{relation_charges 2.4}
\end{equation}
Using this relation, the central charges found here, ($c^*$, $\kappa^*$, $k^*$), can be related to those found in (\ref{central_charges 2.3}), ($c$, $\kappa$, $k$). Explicitly, we have
\begin{equation}
    c^* = 0 - 24 \left(\frac{-L^2}{(16 \pi)^2 G^2} \right) \frac{16 \pi ^2 G}{L^2} = \frac{3}{2G}
\end{equation}
and the relations for $\kappa ^*$ and $k^*$ are trivial. 
Note that $c^*$ is recognized as the Brown-Henneaux central charge for AdS$_3$ gravity \cite{Brown:1986nw}.

From a holographic perspective, the redefiniton of the charges, equation \eqref{changeinalgebra}, corresponds to twisting the stress tensor in the boundary theory. This is the boundary counterpart of performing the change of coordinates from Eddington-Finkelstein-like coordinates to Schwarzschild-like coordinates in the bulk.

\section{Extremal Kerr}\label{sec:extKerr}

\subsection{Geometry and NHEK}\label{}
The analysis of the previous sections can also be applied to extremal Kerr black holes. The metric of the extremal Kerr black hole in Boyer-Lindquist coordinates reads
\begin{equation}
    ds^2 = -\frac{\Delta}{\rho ^2} (dt - a \sin ^2 \theta d\phi )^2 + \frac{\sin ^2 \theta}{\rho ^2} ((r^2 + a^2 ) d\phi - a dt )^2 + \frac{\rho ^2}{\Delta} dr^2 + \rho ^2 d\theta ^2\,,
\end{equation}
where
\begin{equation}
    \Delta = (r - a )^2\, , \quad \rho ^2 = r^2 + a^2 \cos ^2 \theta\, , \quad a = G M \,.
\end{equation}
%In order to study the near-horizon geometry, 
We consider the change of coordinates
\begin{equation}
    \hat{r} = \frac{r - G M}{\lambda G M}\,, \quad \hat{t} = \frac{\lambda t}{2G M}\,, \quad \hat{\phi} = \phi - \frac{t}{2G M}
\end{equation}
and take the limit $\lambda \rightarrow 0$, yielding the near-horizon extremal Kerr (NHEK) geometry
\begin{equation}
\label{nhek 3.4}
    ds^2 = G^2 M^2 (1+ \cos ^2 \theta ) \left( \frac{d\hat{r}^2}{\hat{r}^2} + d\theta ^2 - \hat{r}^2 d\hat{t}^2 \right) + \frac{4 G^2 M^2 \sin ^2 \theta }{1 + \cos ^2 \theta} (d\hat{\phi} + \hat{r} d\hat{t} )^2 .
\end{equation}
Hereafter, we will omit "\textasciicircum " of the coordinates. %, keeping in mind that we are now working in the near-horizon limit. 
In order to apply Godet-Marteau boundary conditions to this metric, we write it in a system
of coordinates similar to the Bondi coordinates
\begin{equation}
\label{BondilikeKerr}
    t = u - \frac{1}{r} \,, \quad \phi = \varphi - \ln r \,,
\end{equation}
such that the metric becomes
\begin{equation}
    ds^2  = G^2 M^2 (1 + \cos ^2 \theta ) (-r^2 du^2 - 2 du dr + d\theta ^2 ) + \frac{4 G^2 M^2 \sin ^2 \theta }{1 + \cos ^2 \theta } (d\varphi + r du )^2 . \label{nhek_bondi 3.1}
\end{equation}

\subsection{Phase Space and Asymptotic Killing Vectors}\label{}
Inspired by the Godet-Marteau boundary conditions for AdS$_2$ \cite{Godet:2020xpk}, we consider the following family of metrics
\begin{align}
    ds^2 &=  G^2 M^2 (1 + \cos ^2 \theta ) ((-r^2 du^2 + 2 P(u) r + 2 T(u) ) du^2 - 2du dr + d\theta ^2) \nonumber \\
    &~~~ + \frac{4G^2 M^2 \sin ^2 \theta }{1+ \cos ^2 \theta} (d\varphi + r du )^2\,, \label{bc_nhek 3.2}
\end{align}
where $P$ and $T$ are arbitrary functions of $u$. They can be obtained from (\ref{nhek_bondi 3.1}) by applying a finite coordinate transformation given by
\begin{equation}
    u \rightarrow \mathcal{F} (u) \,, \quad r \rightarrow \frac{1}{\mathcal{F}'} (r + \mathcal{G}' (u) ) \,, \quad \varphi \rightarrow \varphi - \mathcal{G} (u) \,, \quad \theta \rightarrow \theta \,. \label{finite_transf 3.2}
\end{equation}
The functions $P, T, \mathcal{F}$ and $\mathcal{G}$ are related by 
\begin{align}
    P(u) & = -\mathcal{G}' (u) + \frac{\mathcal{F}''(u)}{\mathcal{F}'(u)}\,, \\
    T(u) & = -\frac12 \mathcal{G}' (u)^2 + \mathcal{G}' (u) \frac{\mathcal{F}''(u)}{\mathcal{F}'(u)} - \mathcal{G}''(u)\,.
\end{align}
The asymptotic Killing vectors generating the transformations (\ref{finite_transf 3.2}) are given by
\begin{equation}
    \xi = \epsilon (u) \partial _u - (r \epsilon ' (u) + \zeta ' (u) ) \partial _r + \zeta (u) \partial _\varphi \,, \label{akv 3.2}
\end{equation}
where $\epsilon (u)$ and $\zeta (u)$ are two arbitrary functions of $u$. By applying the Lie derivative on the metric (\ref{bc_nhek 3.2}), we can also find
\begin{align}
    \delta _\xi P & = \epsilon P' + \epsilon ' P + \epsilon '' + \zeta ' , \\
    \delta _\xi T & = \epsilon T' + 2 \epsilon ' T - \zeta ' P + \zeta '' .
\end{align}

From now on we assume that $u$ is periodic with period $L$. We define the modes
\begin{equation}
    l_n = \xi \left( \frac{L}{2\pi} e^{2\pi i n u /L}, 0 \right), \quad j_n = \xi \left(0 , \frac{L}{2\pi i} e^{2\pi i n u /L} \right),
\end{equation}
where $n \in \mathbb{Z}$. Under the Lie bracket, these modes satisfy the warped Witt algebra \eqref{warpedWitt}.

\subsection{Charge Algebra}\label{}
We can now compute the surface charges by using the expression (\ref{charge_diff 2.3}). For this, we integrate over
$u$ and $\theta$ while keeping $\varphi$ fixed and taking $r \rightarrow \infty$. Defining
\begin{equation}
    h_{\mu \nu} \equiv \delta g_{\mu \nu} = \frac{\partial g_{\mu \nu}}{\partial P} \delta P + \frac{\partial g_{\mu \nu}}{\partial T} \delta T\,,
\end{equation}
we compute
\begin{equation}
    \delta Q_\xi = \frac{G^2 M^2}{4\pi G} \int ^{L} _0 du \int ^\pi _0 d\theta \sin \theta (\delta T \,\epsilon - \delta P \,\zeta )\,, 
\end{equation}
which upon integration yields
\begin{equation}
    Q_\xi = \frac{G M^2}{2\pi} \int ^{L} _0 du (T(u) \epsilon (u) - P(u) \zeta (u) )\,,
\end{equation}
where the NHEK geometry, which has $P(u) = T(u) = 0$, has been chosen as the background metric. In particular, we define
\begin{align}
    & L_n = Q_{l_n} = \frac{G M^2}{2\pi} \int ^{L} _0 du T(u) \frac{L}{2\pi} e^{2\pi i n u / L}, \\
    & J_n = Q_{j_n} = - \frac{G M^2}{2\pi} \int ^{L} _0 du P(u) \frac{L}{2\pi i} e^{2\pi i n u / L}.
\end{align}
The charges $L_n$ and $J_n$ respect the algebra \eqref{WCFTalgebra} with central charges given by
\begin{equation}
    c=0\,, \quad \kappa = \frac{L G M^2}{2 \pi i}\,, \quad k = \frac{L^2 G M^2}{2 \pi^2}\,. \label{central_charges 3.3}
\end{equation}

\subsection{Boundary Conditions in Boyer-Lindquist Coordinates}\label{}
So far, we studied the NHEK geometry 
by writing it in a new system of coordinates (with a retarded time $u$). 
Now, we perform the same analysis in Boyer-Lindquist coordinates. 
Again, we obtain a phase space of metrics from \eqref{nhek 3.4}
by applying the finite coordinate transformation 
\begin{equation}
    t \rightarrow \mathcal{F} (t) \,, \quad r \rightarrow \frac{1}{\mathcal{F}'} (r + \mathcal{G}' (t))\,, \quad \phi \rightarrow \phi - \mathcal{G} (t)\,. \label{finite_transf 3.4}
\end{equation}
Defining $\mathcal{H} (t) \equiv \mathcal{F}''(t) / \mathcal{F}'(t)$, yields
\begin{subequations}
\label{phasespaceBoyerLind}
\begin{align}
    g_{tt} = & \frac{4 r^2 G^2 M^2 \sin ^2 \theta}{1+\cos^2 \theta} - G^2 M^2 (1+\cos ^2 \theta ) (r+\mathcal{G}' (t) )^2 \nonumber \\
    & + \frac{G^2 M^2 (1+\cos ^2 \theta )}{(r+\mathcal{G}' (t) )^2} ((r+\mathcal{G}' (t) ) \mathcal{H} (t) - \mathcal{G}'' (t) )^2 , \label{g_tt 3.4} \\
    g_{tr} = & -G^2 M^2 (1 + \cos ^2 \theta ) \frac{((r+\mathcal{G}' (t) ) \mathcal{H} (t) - \mathcal{G}'' (t))}{(r+\mathcal{G}' (t))^2}\,,\\
    g_{t \theta} = & 0\,, \quad g_{t \phi} = \frac{4 r G^2 M^2 \sin ^2 \theta}{1+ \cos^2 \theta}\,, \\
    g_{rr} = & \frac{G^2 M^2 (1 + \cos ^2 \theta )}{(r+\mathcal{G}' (t))^2}\,, \quad g_{r \theta} = 0\,, \quad g_{r \phi} = 0\,, \\
    g_{\theta \theta} = & G^2 M^2 (1+ \cos ^2 \theta ) \,, \quad g_{\theta \phi} = 0\,, \quad g_{\phi \phi} = \frac{4 G^2 M^2 \sin ^2 \theta}{1 + \cos^2 \theta}\,, \label{g_phi_phi 3.4}
\end{align}
\end{subequations}
where the NHEK geometry is obtained by setting $\mathcal{G}' (t) = 0$ and $\mathcal{H} (t) = 0$. Hence, the order of the non-zero fluctuations of the boundary metric is given by 
\begin{equation}
\label{fluctuationskerr}
    h_{tt} = \mathcal{O} (r)\,, \quad h_{tr} = \mathcal{O} (r^{-1})\,, \quad h_{rr} = \mathcal{O} (r^{-3})\,.
\end{equation}
The asymptotic Killing vectors generating the transformations (\ref{finite_transf 3.4}) are given by
\begin{equation}
    \xi (\epsilon , \zeta ) = \epsilon (t) \partial _t + (-r \epsilon ' (t) - \zeta ' (t)) \partial _r + \zeta (t) \partial _\phi \,,\label{akv 3.4}
\end{equation}
where $\epsilon (t)$ and $\zeta (t)$ are two arbitrary functions of $t$. 

We recall that the group of exact isometries of the NHEK geometry, $SL(2,\mathbb{R}) \times U(1)$, is generated by the Killing vectors
\begin{align}
    & \xi _{-1} = \partial _t \,, \quad \xi _0 = t \partial _t -r \partial _r \,, \quad \xi _1 = \left( t^2 + \frac{1}{r^2} \right) \partial _t - 2 t r \partial _r - \frac{2}{r} \partial _\phi \,, \label{isometries_a 3.4} \\
    & \xi _\phi = \partial _\phi \,. \label{isometries_b 3.4}
\end{align}
Comparing these vectors with (\ref{akv 3.4}), we find that $\xi _{-1} = \xi (\epsilon =1, \zeta =0)$, $\xi _0 = \xi (\epsilon =t , \zeta =0)$, $\xi _\phi = \xi (\epsilon =0, \zeta =1)$ and that $\xi _1$ correspond to $\xi (\epsilon =t^2 , \zeta =0)$ up to subleading terms in $r$. Hence, the asymptotic symmetry group contains all the exact isometries of the NHEK geometry, which was not the case for the boundary conditions studied in \cite{Guica:2008mu}.

By applying the Lie derivative on \eqref{phasespaceBoyerLind}, we find 
\begin{align}
    & \delta _{\xi} \mathcal{G}' (t) = \epsilon '(t) \mathcal{G}' (t) + \epsilon (t) \mathcal{G}'' (t) - \zeta '(t) \,, \\
    & \delta _{\xi} \mathcal{H} (t) = \epsilon '(t) \mathcal{H} (t) + \epsilon '' (t) + \epsilon (t) \mathcal{H} '(t)\,.
\end{align}
From now on we assume that $t$ is periodic with period $L$. We define modes as
\begin{equation}
    l_n = \xi \left(\frac{L}{2\pi} e^{2\pi i n t /L} , 0\right) , \quad j_n = \xi \left( 0, \frac{L}{2\pi i} e^{2\pi i n t /L} \right),
\end{equation}
with $n \in \mathbb{Z}$, which satisfy \eqref{warpedWitt}.

Here, the integral considered in (\ref{charge_diff 2.3}) for the computation of the charges is taken over $t$ and $\theta$ while $r \rightarrow \infty$ and $\phi$ is constant. 
Computing the charges explicitly, we find
\begin{equation}
    Q_\xi = \frac{G M^2}{4\pi } \int _0 ^{L} dt (2 \mathcal{G}' (t) \zeta (t) - \mathcal{G}' (t) ^2 \epsilon (t) + 2 \epsilon ' (t) \mathcal{H} (t) + \epsilon (t) \mathcal{H} (t)^2 )\,,
\end{equation}
where the NHEK geometry, which has $\mathcal{G}' (t) = \mathcal{H} (t) = 0$, has been chosen as the background metric. In particular, we define
\begin{align}
    & L_n = Q_{l_n} = \frac{G M^2 }{4\pi} \int _0 ^{L} dt \left( -\mathcal{G}' (t) ^2 + \frac{4\pi i n}{L} \mathcal{H} (t) + \mathcal{H} (t) ^2 \right) \frac{L}{2\pi} e^{2\pi i n t /L}, \\
    & J_n = Q_{j_n} = \frac{G M^2}{4\pi} \int _0 ^{L} dt \; 2 \; \mathcal{G}'(t) \,\frac{L}{2\pi i} e^{2\pi i n t /L}.
\end{align}
The charges $L_n$ and $J_n$ fulfill the algebra \eqref{WCFTalgebra} with central charges given by
\begin{equation}
\label{Kerrcstarkstar}
    c^* = 12 G M^2 = 12 J\,, \quad \kappa ^* = 0 \,, \quad k^* = \frac{L^2 G M^2}{2 \pi^2} = \frac{J L^2 }{2\pi^2} \,.
\end{equation}

 The algebra \eqref{WCFTalgebra} with central charges \eqref{Kerrcstarkstar}, $(c^*, \kappa^*,k^*)$ can be related to the one with central charges (\ref{central_charges 3.3}), $(c, \kappa,k)$,  by the transformation \eqref{changeinalgebra} and \eqref{relation_charges 2.4}.
 Indeed, we have
\begin{equation}
    c^* = 0 - 24 \left( \frac{L G M^2}{2 \pi i} \right)^2 \frac{2\pi^2}{L^2 G M^2} = 12 G M^2
\end{equation}
and the relations for $\kappa ^*$ and $k^*$ are trivial. 
Here $c^*$ is recognized as the Kerr/CFT central charge \cite{Guica:2008mu}.

\subsection{Comparison to other Boundary Conditions for extremal Kerr Black Holes}\label{sec:comp}
We now compare our results with those obtained in \cite{Matsuo:2009sj}. There, the perturbations defined on the background metric (\ref{nhek 3.4}) were 
\begin{subequations}
\label{Matsuobc}
\begin{align}
     h_{tt} &= \mathcal{O} (r^0)\,,& h_{tr} &= \mathcal{O} (r^{-3})\,, & h_{t \theta} &= \mathcal{O} (r^{-3})\,,  &  h_{t \phi} &= \mathcal{O} (r^{-2})\,, &\\
    h_{rr} &= \mathcal{O} (r^{-4})\,,  & h_{r \theta} &= \mathcal{O} (r^{-4})\,, & h_{r \phi} &= \mathcal{O} (r^{-3})\,, & \\
     h_{\theta \theta} &= \mathcal{O} (r^{-3})\,, & h_{\theta \phi} &= \mathcal{O} (r^{-3})\,, & h_{\phi \phi} &= \mathcal{O} (r^{-2})
\end{align}
\end{subequations}
and the vectors solving the asymptotic Killing equation took the general form
\begin{align}
    \xi = & \left( \epsilon (t) + \frac{\epsilon '' (t)}{2 r^2} + \mathcal{O} (r^{-3}) \right) \partial _t + \left( -r \epsilon ' (t) + \frac{\epsilon ''' (t)}{2r} + \mathcal{O} (r^{-2}) \right) \partial _r \nonumber \\
    & + \left( \mathcal{C} - \frac{ \epsilon '' (t)}{r} + \mathcal{O} (r^{-3}) \right) \partial _\phi + \mathcal{O} (r^{-3}) \partial _\theta \,, \label{akv 3.5}
\end{align}
where $\epsilon (t)$ is an arbitrary function of $t$ and $\mathcal{C}$ is an arbitrary constant. The boundary conditions \eqref{Matsuobc} are different from ours, compare equation \eqref{fluctuationskerr}. Neglecting the subleading terms, we see that (\ref{akv 3.4}) reduces to \eqref{akv 3.5} upon setting $\zeta(t) = \mathcal{C} = \mathrm{const.}$ 
Hence, in both cases the expression (\ref{akv 3.5}) contains the vectors (\ref{isometries_a 3.4})-(\ref{isometries_b 3.4}) generating the $SL(2,\mathbb{R}) \times U(1)$ group of isometries.

In \cite{Matsuo:2009sj} it is claimed that the charges associated to the vectors (\ref{akv 3.5}) with $\mathcal{C} = 0$ form a Virasoro algebra with vanishing central extension, contrary to our result. 
Indeed, restricting to a subset of our charges by considering only asymptotic Killing vectors (\ref{akv 3.4}) that have $\zeta (t) = 0$,
we obtain a Virasoro algebra (\ref{WCFT_LL 2.3}) with non-zero central charge. 

Different boundary conditions encompassing the NHEK geometry were also presented in \cite{Kapec:2019hro, Castro:2019crn}. Starting from the background metric (\ref{nhek 3.4}), 
a phase space of metrics was obtained by applying a finite coordinate transformation
\begin{align}
    & t \rightarrow f(t) + \frac{2 f''(t) f'(t)^2}{4 r^2 f'(t)^2 - f''(t)^2} \; , \nonumber \\
    & r \rightarrow \frac{4 r^2 f'(t)^2 - f''(t)^2}{4 r f'(t)^3} \; , \label{finite_transf 3.5} \\
    & \phi \rightarrow \phi + \log \left( \frac{2 r f'(t) - f''(t)}{2 r f'(t) + f''(t)} \right)\,, \nonumber
\end{align}
yielding the line element
\begin{align}
    ds ^2 = & G^2 M^2 (1 + \cos ^2 \theta ) \left( -r^2 \left( 1 + \frac{\{ f(t) , t\}}{2 r^2} \right)^2 dt^2 + \frac{dr^2}{r^2} + d\theta ^2 \right) \nonumber \\
    & + \frac{4 G^2 M^2 \sin ^2 \theta}{1 + \cos ^2 \theta} \left( d \phi + r \left( 1 - \frac{\{ f(t), t \}}{2r^2} \right) dt \right)^2 
\end{align}
with the Schwarzian derivative
\begin{equation}
    \{ f(t) , t \} = \left( \frac{f''}{f'} \right)' - \frac12 \left( \frac{f''}{f'}\right) ^2 .
\end{equation}
Equivalently, the components of this metric read
\begin{subequations}
\begin{align}
    g_{tt} = & \frac{4 r^2 G^2 M^2 \sin ^2 \theta}{1 + \cos ^2 \theta} \left( 1 - \frac{\{ f(t), t \}}{2r^2} \right)^2  - G^2 M^2 (1 + \cos ^2 \theta ) r^2 \left( 1 + \frac{\{ f(t) , t\}}{2 r^2} \right)^2 , \\
    g_{tr} = & 0\,, \quad  g_{t \theta} = 0\,, \quad g_{t \phi} = \frac{4 G^2 M^2 \sin ^2 \theta}{1 + \cos ^2 \theta} r \left( 1 - \frac{\{ f(t), t \}}{2r^2} \right),\\
    g_{rr} = & \frac{G^2 M^2 (1 + \cos ^2 \theta )}{r^2} \,, \quad g_{r \theta} = 0\,, \quad g_{r \phi} = 0\,, \\
    g_{\theta \theta} = & G^2 M^2 (1 + \cos ^2 \theta )\,, \quad g_{\theta \phi} = 0\,, \quad g_{\phi \phi } = \frac{4 G^2 M^2 \sin ^2 \theta}{1 + \cos ^2 \theta}\,,
\end{align}
\end{subequations}
which are different from the components \eqref{phasespaceBoyerLind} that we obtained from applying the transformation (\ref{finite_transf 3.4}). The order of the non-zero fluctuations of the boundary metric 
\begin{equation}
\label{fluctuationskerr2}
    h_{tt} = \mathcal{O} (r^{-2})\,, \quad h_{t\varphi} = \mathcal{O} (r^{-1})
\end{equation}
are different from  \eqref{Matsuobc} and ours, compare equation \eqref{fluctuationskerr}.
Furthermore, while here the components only depend on one free function of $t$,  our class of metrics (\ref{g_tt 3.4})-(\ref{g_phi_phi 3.4}) depends on two. 
Expanding $f(t) = t + \epsilon (t) + \mathcal{O} (\epsilon ^2)$, the asymptotic Killing vectors generating the transformations (\ref{finite_transf 3.5}) are given by
\begin{equation}
    \xi = \left( \epsilon (t) + \frac{\epsilon '' (t)}{2 r^2} \right) \partial _t - r \epsilon ' (t) \partial _r - \frac{\epsilon '' (t)}{r} \partial _\phi \,,
\end{equation}
where $\epsilon (t)$ is an arbitrary function of $t$. Again, up to subleading terms, these vectors are a subset of the vectors (\ref{akv 3.4}), obtained by setting $\zeta (t) = 0$.

\section{Ultra-cold Kerr-dS}\label{sec:extKerrdS}

\subsection{Geometry and Phase Space}\label{}

In this section, we study the near-horizon geometry of the Kerr-dS black hole in the ultracold limit where
the inner, outer and cosmological horizon coincide. In this limit, the metric takes the form \cite{Anninos:2009yc} 
\begin{equation}
\frac{ds^2}{\ell^2}= \Gamma(\theta)\Big(-d t^2+d r^2+ \alpha(\theta)d\theta^2\Big)+\gamma(\theta)(d\phi + {\bar k} r d t)^2
\end{equation}
with 
\begin{align}
 \Gamma(\theta) &=   \frac{\sqrt{2 \sqrt{3}-3} \left(\left(3-2 \sqrt{3}\right)
   \cos ^2(\theta )-1\right)}{2 \left(\sqrt{3}-3\right)}\,, &
   \alpha(\theta) &= \frac{2 \sqrt{14 \sqrt{3}-24} \,  }{\left(7 \sqrt{3}-12\right)
   \cos ^2(\theta )+\sqrt{3}}\,, \\
  \gamma(\theta) &=  \frac{\sin ^2(\theta ) \left(\left(15 \sqrt{3}-26\right) \cos
   ^2(\theta )+\sqrt{3}-2\right)}{3 \left(4 \sqrt{3}-7\right)
   \cos (2 \theta )+8 \sqrt{3}-15} \,, &
   {\bar k} &= -\sqrt{3}\,,
\end{align}
where the bar has been introduced to avoid possible confusions between the parameter $\bar k$ with the central extension $k$.
Here, we have chosen our units such that the cosmological constant $\Lambda = 3/\ell^2$, with $\ell$ being the dS radius. The sign of $\bar k$ is arbitrary and can be changed by sending $t \to -  t$.
We change to Eddington-Finkelstein-like coordinates 
\begin{equation}
u=t - r\,, \quad \phi = \bar \varphi - \frac{\bar k r^2}{2}\,,
\end{equation}
yielding
\begin{equation}
\frac{ds^2}{\ell^2}= \Gamma(\theta)(- du^2 - 2dudr+\alpha(\theta)d\theta^2)+\gamma(\theta)(d \bar \varphi+ \bar k rdu)^2\,.
\label{UCmetric}
\end{equation}
Upon setting 
\begin{equation}
   \bar \varphi = \bar k \varphi\,, \qquad \gamma(\theta)= \frac{\bar \gamma(\theta)}{{\bar k}^2}\,,
\end{equation}
we get
\begin{equation}
\label{UCmetrick1}
    \frac{ds^2}{\ell^2}= \Gamma(\theta)(- du^2 - 2dudr+\alpha(\theta)d\theta^2)+\bar\gamma(\theta)(d \varphi+ rdu)^2\,.
\end{equation}
Inspired by \cite{Afshar:2019axx}, we consider the following family of metrics
\begin{equation}
\frac{ds^2}{\ell^2}=\Gamma(\theta)\Big(2\big(P(u)r+T(u)\big)du^2 - 2dudr+ \alpha(\theta) d\theta^2\Big)+ \bar{\gamma}(\theta)(d\varphi+ rdu)^2\,,
\label{uplift_UC}
\end{equation}
where $P$ and $T$ are arbitrary functions of $u$.
This family can be obtained by applying the finite coordinate transformation 
\begin{equation}
 u \rightarrow \mathcal{F}(u), \quad r \rightarrow \frac{1}{\mathcal{F}'} (r + \mathcal{G}' (u)), \quad \varphi \rightarrow \varphi - \mathcal{G} (u)
 \label{finite_trafo_uplift}
 \end{equation}
 to \eqref{UCmetrick1}.
The functions $P, T, \mathcal{F}$ and $\mathcal{G}$ are related by
\begin{equation}
T(u) = - \frac{1}{2} \mathcal{F}'(u)^2 - \mathcal{G}''(u)+\frac{\mathcal{G}'(u)\mathcal{F}''(u)}{\mathcal{F}'(u)}\,, \quad
P(u) = \frac{\mathcal{F}''(u)}{\mathcal{F}'(u)}\,.
\end{equation}

\subsection{Asymptotic Killing Vectors}\label{}
The asymptotic Killing vectors generating the transformations \eqref{finite_trafo_uplift} read
\begin{equation}
\xi =\epsilon(u)\partial_u -(r\epsilon'(u)+\zeta'(u))\partial_r+\zeta(u)\partial_{\varphi}\,,
\label{diff_UC}
\end{equation}
where  $\epsilon(u)$ and $\zeta(u)$ are two arbitrary functions of $u$. 
We take the retarded time $u$ to be periodic 
with period $L$, and define the generators 
\begin{equation}
l_n =  \xi (\epsilon =  \frac{L}{2 \pi} e^{2\pi i n u /L}, \zeta = 0)\,,\quad j_n =  \xi (\epsilon = 0,
\zeta = \frac{L}{2\pi i}e^{2\pi i n u /L})\,,
\label{UC_generators}
\end{equation}
which obey \eqref{warpedWitt}.
By applying the Lie derivative on the metric \eqref{uplift_UC}, we find the variations of $T(u)$ and $P(u)$ 
\begin{subequations}
\begin{align}
    \delta_{\xi} T(u) &=  \Big(2T(u) \epsilon'(u) +\epsilon(u)T'(u)-P(u)\zeta'(u) +\zeta''(u)\Big)\,,  \\
    \delta_{\xi} P(u) &= \Big(P(u)\epsilon'(u)+\epsilon(u)P'(u)+\epsilon''(u)\Big).
\end{align}
\end{subequations}

\subsection{Charge Algebra}\label{}
We compute the surface charges from (\ref{charge_diff 2.3}), yielding
\begin{equation}
Q= \frac{\ell^2}{8\pi  G}\int_0^{L}du (\sqrt{3}-1)(\epsilon(u) T(u) -\zeta(u) P(u))\,,
\label{UC_charges}
\end{equation}
where we have integrated over a constant $r,\varphi$ surface and taken the limit $r \to \infty$. 
Defining
\begin{subequations}
\begin{align}
   L_n=Q_{l_n} &= \frac{L\, \ell^2}{16 \pi^2 G} \int_0^{L}du (\sqrt{3}-1)  e^{2\pi i n u /L} T(u)\,, \\
    J_n=Q_{j_n} &= -\frac{L\, \ell^2}{16 \pi^2 i  G}\int_0^{L}du (\sqrt{3}-1) e^{2\pi i n u /L} P(u)\,,
\end{align}
\end{subequations}
 one readily computes that the charges $L_n, J_n$ obey \eqref{WCFTalgebra} with $c = k = 0$ and 
 \begin{equation}
     \kappa  = \frac{1}{i} \frac{L \, \ell^2}{8 \pi G} \left(\sqrt{3}-1\right)\,.
 \end{equation}

\section{Conclusion}
\label{sec:conclusion}
\noindent
In this paper, we studied new boundary conditions for the near-horizon geometries of extremal black holes in three and four dimensions. Our boundary conditions for extremal BTZ and Kerr black holes were obtained by uplifting the boundary conditions by Godet and Marteau \cite{Godet:2020xpk}, from two to three or from two to four dimensions. 
In the case of the ultra-cold Kerr-dS black hole, our boundary conditions were obtained by uplifting the boundary conditions by Afshar, Gonz\'alez, Grumiller and Vassilevich \cite{Afshar:2019axx} from two to four dimensions.
This shows that certain boundary conditions existing in 2d gravity can be uplifted to higher dimensions in a natural way. 

We studied the asymptotic symmetries preserving these boundary conditions and associated charges. 
Our charges are computed by integrating over time on a constant azimuthal angle surface, instead of doing it vice versa (integrating over the azimuthal angle on a constant time surface).
To obtain finite charges, Lorentzian time necessarily needs to be periodic --- this is to be understood as the Wick rotation of the periodicity in Euclidean time. Switching angular and temporal circle for the computation of the charges is motivated 
from modular invariance of the putative dual field theory, as detailed in the introduction. This introduces a time scale 
$L = i \beta$ in our charges and central extensions, where $\beta$ is the temperature.
%\textcolor{blue}{While this dependence on $L$ may seem surprising at first, we note that also }
In this way, we obtain non-trivial charges which span a Virasoro-Kac-Moody algebra, the symmetry algebra of a warped conformal field theory. The results for the central extensions are summarized in table \ref{table:1}. 
For the case of the extremal BTZ and Kerr black holes we studied boundary conditions and the associated asymptotic symmetry algebras in two different systems of coordinates. A priori, such boundary conditions are not equivalent, as when it comes to asymptotic symmetries and charges, diffeomorphisms can have non-vanishing associated charges and thus carry non-trivial information.
Having different boundary conditions available, the choice of boundary conditions is related to how the asymptotic boundary of spacetime is approached --- in our case: following spacelike curves in Schwarzschild-like coordinates or null curves in Eddington-Finkelstein-like coordinates. 
However, even if one fixes the direction of approach to the asymptotic boundary, different boundary conditions are possible. The particular choice of boundary conditions is not unique and depends on the physical situation at hand. 
%This is demonstrated by the example of asymptotically AdS$_3$ spacetimes, where in addition to the pioneering analysis of \cite{Brown:1986nw}, distinct boundary conditions were developed in \cite{Compere:2013bya} and later considered in \cite{Grumiller:2016pqb}.
Our analysis for extremal BTZ and Kerr black holes yields, in each case a Virasoro-Kac-Moody algebra, albeit with different central extensions. We then showed that in both cases these different central extensions can be related by a mere redefinition of generators, showing that the two algebras are isomorphic.
 This mirrors the analysis of \cite{Barnich:2012aw}, which studied the asymptotic symmetries of three-dimensional asymptotically AdS spacetime in Bondi gauge, yielding a Virasoro algebra, the algebra found in the previous analysis performed in the Fefferman-Graham gauge \cite{Brown:1986nw}.
In the case of extremal Kerr black holes we relate our results to boundary conditions which have previously been studied in the literature \cite{Matsuo:2009sj, Kapec:2019hro, Castro:2019crn}. For this, we have to truncate our asymptotic symmetries to a Virasoro $\oplus$ $u(1)$ algebra.  Contrary to \cite{Matsuo:2009sj}, we find, that the Virasoro algebra has non-vanishing central charge, paving the way for possible microstate countings using asymptotic symmetries, in the spirit of \cite{Guica:2008mu}.
%In particular, the values obtained for the central extensions were shown to contradict the result claimed in \cite{Matsuo:2009sj}. 
\\
Lastly, we studied the near-horizon geometry of the ultra-cold Kerr-dS black hole whose holographic interpretation has so far been elusive. We find, using an uplift of the  boundary conditions \cite{Afshar:2019axx}, that the asymptotic symmetries span a Virasoro-Kac-Moody algebra, thereby providing first evidence that warped conformal field theories could be  the holographic dual for such black holes.  

While our analysis is purely classical, our results suggest that warped conformal field theories provide a holographic description of extremal black holes. This kinematical observation, based on symmetries, could be pushed in various directions, such as entropy matchings and perturbation theory to put the proposal on firmer grounds.
In line with this, the question arises, whether the quantum theories obtained by performing standard canonical quantization are unitary. In the case of the extremal Kerr and BTZ black holes, due to the isomorphism mentioned above,  it suffices to consider whether representation with central charges $(c, 0, k)$ can be unitary. We answer this question in the negative, as in our case $c > 0$ and $k <0$, c.f. table \ref{table:1}, due to  $L \in i \mathbb{R}$ and $M \in \mathbb{R}$. However, to have unitary highest-weight representation it is necessary to have $c > 0$ and $k >0$, see \cite[Section 2.3]{Detournay:2012pc}. For the case of the ultracold Kerr-dS black hole, the answer is not clear, since we cannot make the redefinition and representations with $k = 0$ and $\kappa \neq 0$ have not been discussed in the literature to the best of our knowledge.

In the context of holography, WCFTs with a positive central charge but a negative $U(1)$ level have appeared generically. Despite featuring negative norm descendant states that violate unitarity, some of their properties are kept under good control. For instance, it was shown that the modular bootstrap remains feasible in theories with mild violations of unitarity, where the negative norm states can be resummed into a Virasoro-Kac-Moody character whose contribution to the bootstrap equations is positive \cite{Apolo:2018eky}. In fact, any WCFT with a negative level must feature at least two states with imaginary U(1) charge, rendering the Hamiltonian non-hermitian. However, that feature is essential for the WCFT counterpart of the Cardy formula to be able to reproduce the entropy of WAdS$_3$ black holes \cite{Detournay:2012pc}. Furthermore, the study of the extremal limits of WAdS$_3$ black holes and WCFTs (in the spirit of \cite{Ghosh:2019rcj} for 2d CFTs) has revealed the emergence of a universal Schwarzian sector (as expected on general grounds for extremal black holes \cite{Maldacena:2016upp, Iliesiu:2020qvm}), but only when the seed theory was non-unitary \cite{Aggarwal:2022xfd, Aggarwal:2023peg}.  

We leave it to future work to exploit our results to  establish a potential dual holographic description of extremal black holes in terms of a warped conformal field theory and study the microscopic description of extremal black holes.
%In section two, we proposed boundary conditions for extremal BTZ black holes, both in Eddington-Finkelstein coordinates, as well as in Schwarzschild coordinates. We calculated the asymptotic symmetries and showed that they the charges span a Virasoro-Kac-Moody algebra in both cases, the symmetry algebra of a warped conformal field theory, albeit with different central charges. We then showed that these central extensions could be related by a redefinition of generators.
%some boundary conditions from two-dimensional gravity in order to obtain new boundary conditions for higher-dimensional black holes.
%In the different cases studied, the charges associated to the asymptotic symmetries were shown to satisfy the symmetry algebra of a WCFT. 
%The central extensions obtained are summarized in table \ref{table:1}.
\begin{table}
\centering
\begin{tabular}{ |m{2.2cm}||c|c|c|c| } 
\hline
Black hole & $c$ & $\kappa$ & $k$ & Relation between the central charges \\
\hline
Extremal BTZ $(u, r, \varphi)$ & $0$ & $\frac{L}{16 \pi i G}$ & $\frac{L^2}{16 \pi^2 G}$ & \multirow{2}{3.5cm}{$c^* = c - 24 \kappa ^2 /k \, ,$ \newline $\kappa ^* = 0 \, ,$ \newline $k^* = k$} \\
\cline{1-4}
Extremal BTZ $(\tau , \rho , \varphi )$& $\frac{3}{2G}$ & $0$ & $\frac{L^2}{16 \pi ^2 G}$ & \\
\hline 
Extremal Kerr $(u , r, \theta , \varphi)$& $0$ & $\frac{L G M^2}{2 \pi i}$ & $\frac{L^2 G M^2}{2 \pi^2}$ & \multirow{2}{3.5cm}{$c^* = c - 24 \kappa ^2 /k \, ,$ \newline $\kappa ^* = 0 \, ,$ \newline $k^* = k$} \\ 
\cline{1-4}
Extremal Kerr $(t , r, \theta , \phi)$& $12 G M^2$ & $0$ & $\frac{L^2 G M^2}{2 \pi^2}$ & \\
\hline
Ultra-cold Kerr-dS $(u , r, \theta , \varphi)$& $0$ & $\frac{L \, \ell^2\left(\sqrt{3}-1\right)}{8 \pi i G}$ & $0$ & \cellcolor{lightgray} \\
\hline
\end{tabular}
\caption{Central charges obtained for different black holes in different systems of coordinates: the central charges, $(c, \kappa, k)$, found by studying the asymptotic symmetries in Eddington-Finkelstein coordinates are related to the central charges, $(c^*, \kappa^*, k^*)$, found by studying the asymptotic symmetries in Schwarzschild-like coordinates due to the isomorphism \eqref{changeinalgebra}.}
\label{table:1}
\end{table}

\noindent

%\textcolor{blue}{For the near-horizon geometry of the extremal BTZ black hole, the boundary conditions could be applied in two different systems of coordinates. The central extensions were different in either case but it was shown that they could be related by a mere redefinition of the generators. A similar result was derived for the near-horizon geometry of the extremal Kerr black hole. In that case, a comparison was also made with other boundary conditions proposed in the past. In particular, the values obtained for the central extensions were shown to contradict the result claimed in \cite{Matsuo:2009sj}. Another case studied in this paper was the near-horizon geometry of the Kerr-dS black hole in the ultracold limit. The fact that the charges satisfy a WCFT algebra provides an interesting insight into the building of a holographic dual for this theory.}

\section*{Acknowledgements}

The authors thank Dionysios Anninos, Alejandra Castro, Daniel Grumiller, Tom Hartman, and Chiara Toldo for useful discussions and exchanges on the topics covered in this work. We thank Katharina Schäfer for initial collaboration on Sect. 4 of this paper. RW acknowledges support of the Fonds de la Recherche Scientifique F.R.S.-FNRS (Belgium) through 
the PDR/OL C62/5 project ``Black hole horizons: away from conformality'' (2022-2025) and RW thanks the Erwin Schrödinger Institute for hospitality, where part of this work was carried out. 
RW also acknowledges support by the
Heising-Simons Foundation under the “Observational Signatures of Quantum Gravity” collaboration grant 2021-2818 and the U.S. Department of Energy, Office of High Energy Physics,
under Award No. DE-SC0019470 during the final stages of this work.
TS is a Research Fellow of the Fonds de la Recherche Scientifique F.R.S.– FNRS (Belgium). SD is a Senior Research Associate of the Fonds de la Recherche Scientifique F.R.S.-FNRS (Belgium).
SD was supported in part by IISN – Belgium (convention 4.4503.15) and benefited from the support
of the Solvay Family. SD acknowledges support of the Fonds de la Recherche Scientifique F.R.S.-
FNRS (Belgium) through the CDR project C 60/5 - CDR/OL “Horizon holography: black holes
and field theories” (2020-2022), and the PDR/OL C62/5 project “Black hole horizons: away from
conformality” (2022-2025)


% Provide your bibliography here. You have two options:

%%% FIRST OPTION
% Write your entries here directly, following the example below, including:
% Author(s), Title, Journal Ref. with year in parentheses at the end, followed by the DOI number.

%\begin{thebibliography}{99}
%\bibitem{1931_Bethe_ZP_71} H. A. Bethe, {\it Zur Theorie der Metalle. i. Eigenwerte und Eigenfunktionen der linearen Atomkette}, Zeit. f{\"u}r Phys. {\bf 71}, 205 (1931), \doi{10.1007\%2FBF01341708}.
%\bibitem{arXiv:1108.2700} P. Ginsparg, {\it It was twenty years ago today... }, \url{http://arxiv.org/abs/1108.2700}.
%\end{thebibliography}

% Use your bibtex library, formatted by the SciPost style file.
\bibliography{references.bib}

\begin{thebibliography}{100}
\providecommand{\url}[1]{\texttt{#1}}
\providecommand{\urlprefix}{URL }
\expandafter\ifx\csname urlstyle\endcsname\relax
  \providecommand{\doi}[1]{doi:\discretionary{}{}{}#1}\else
  \providecommand{\doi}{doi:\discretionary{}{}{}\begingroup \urlstyle{rm}\Url}\fi
\providecommand{\eprint}[2][]{\url{#2}}

\bibitem{Fok1959-FOKTTO}
V.~A. Fok,
\newblock \emph{The Theory of Space, Time and Gravitation},
\newblock Macmillan, New York, (1959).

\bibitem{Bunster:2014mua}
C.~Bunster, M.~Henneaux, A.~Perez, D.~Tempo and R.~Troncoso,
\newblock \emph{{Generalized Black Holes in Three-dimensional Spacetime}},
\newblock JHEP \textbf{05}, 031 (2014),
\newblock \doi{10.1007/JHEP05(2014)031},
\newblock \eprint{1404.3305}.

\bibitem{Kostant1970QuantizationAU}
B.~Kostant,
\newblock \emph{Quantization and unitary representations},
\newblock In C.~T. Taam, ed., \emph{Lectures in Modern Analysis and Applications III}, pp. 87--208. Springer Berlin Heidelberg, Berlin, Heidelberg,
\newblock ISBN 978-3-540-36417-7 (1970).

\bibitem{MR0260238}
J.-M. Souriau,
\newblock \emph{Structure des syst\`emes dynamiques},
\newblock Dunod, Paris,
\newblock Ma\^{i}trises de math\'{e}matiques (1970).

\bibitem{Bondi:1962px}
H.~Bondi, M.~G.~J. van~der Burg and A.~W.~K. Metzner,
\newblock \emph{{Gravitational waves in general relativity. 7. Waves from axisymmetric isolated systems}},
\newblock Proc. Roy. Soc. Lond. A \textbf{269}, 21 (1962),
\newblock \doi{10.1098/rspa.1962.0161}.

\bibitem{Sachs:1962wk}
R.~K. Sachs,
\newblock \emph{{Gravitational waves in general relativity. 8. Waves in asymptotically flat space-times}},
\newblock Proc. Roy. Soc. Lond. A \textbf{270}, 103 (1962),
\newblock \doi{10.1098/rspa.1962.0206}.

\bibitem{deBoer:2003vf}
J.~de~Boer and S.~N. Solodukhin,
\newblock \emph{{A Holographic reduction of Minkowski space-time}},
\newblock Nucl. Phys. B \textbf{665}, 545 (2003),
\newblock \doi{10.1016/S0550-3213(03)00494-2},
\newblock \eprint{hep-th/0303006}.

\bibitem{Barnich:2009se}
G.~Barnich and C.~Troessaert,
\newblock \emph{{Symmetries of asymptotically flat 4 dimensional spacetimes at null infinity revisited}},
\newblock Phys. Rev. Lett. \textbf{105}, 111103 (2010),
\newblock \doi{10.1103/PhysRevLett.105.111103},
\newblock \eprint{0909.2617}.

\bibitem{Barnich:2010eb}
G.~Barnich and C.~Troessaert,
\newblock \emph{{Aspects of the BMS/CFT correspondence}},
\newblock JHEP \textbf{05}, 062 (2010),
\newblock \doi{10.1007/JHEP05(2010)062},
\newblock \eprint{1001.1541}.

\bibitem{Campiglia:2014yka}
M.~Campiglia and A.~Laddha,
\newblock \emph{{Asymptotic symmetries and subleading soft graviton theorem}},
\newblock Phys. Rev. D \textbf{90}(12), 124028 (2014),
\newblock \doi{10.1103/PhysRevD.90.124028},
\newblock \eprint{1408.2228}.

\bibitem{Campiglia:2015yka}
M.~Campiglia and A.~Laddha,
\newblock \emph{{New symmetries for the Gravitational S-matrix}},
\newblock JHEP \textbf{04}, 076 (2015),
\newblock \doi{10.1007/JHEP04(2015)076},
\newblock \eprint{1502.02318}.

\bibitem{Strominger:2013jfa}
A.~Strominger,
\newblock \emph{{On BMS Invariance of Gravitational Scattering}},
\newblock JHEP \textbf{07}, 152 (2014),
\newblock \doi{10.1007/JHEP07(2014)152},
\newblock \eprint{1312.2229}.

\bibitem{Strominger:2017zoo}
A.~Strominger,
\newblock \emph{{Lectures on the Infrared Structure of Gravity and Gauge Theory}},
\newblock ISBN 978-0-691-17973-5 (2017), \eprint{1703.05448}.

\bibitem{Brown:1986nw}
J.~D. Brown and M.~Henneaux,
\newblock \emph{{Central Charges in the Canonical Realization of Asymptotic Symmetries: An Example from Three-Dimensional Gravity}},
\newblock Commun.Math.Phys. \textbf{104}, 207 (1986),
\newblock \doi{10.1007/BF01211590}.

\bibitem{Maldacena:1998bw}
J.~M. Maldacena and A.~Strominger,
\newblock \emph{{AdS(3) black holes and a stringy exclusion principle}},
\newblock JHEP \textbf{12}, 005 (1998),
\newblock \doi{10.1088/1126-6708/1998/12/005},
\newblock \eprint{hep-th/9804085}.

\bibitem{Banados:1992gq}
M.~Banados, M.~Henneaux, C.~Teitelboim and J.~Zanelli,
\newblock \emph{{Geometry of the (2+1) black hole}},
\newblock Phys. Rev. \textbf{D48}, 1506 (1993),
\newblock \doi{10.1103/PhysRevD.48.1506, 10.1103/PhysRevD.88.069902},
\newblock [Erratum: Phys. Rev.D88,069902(2013)],
\newblock \eprint{gr-qc/9302012}.

\bibitem{Banados:1992wn}
M.~Banados, C.~Teitelboim and J.~Zanelli,
\newblock \emph{{The Black hole in three-dimensional space-time}},
\newblock Phys. Rev. Lett. \textbf{69}, 1849 (1992),
\newblock \doi{10.1103/PhysRevLett.69.1849},
\newblock \eprint{hep-th/9204099}.

\bibitem{Strominger:1998yg}
A.~Strominger,
\newblock \emph{{AdS(2) quantum gravity and string theory}},
\newblock JHEP \textbf{01}, 007 (1999),
\newblock \doi{10.1088/1126-6708/1999/01/007},
\newblock \eprint{hep-th/9809027}.

\bibitem{Cvetic:1998xh}
M.~Cvetic and F.~Larsen,
\newblock \emph{{Near horizon geometry of rotating black holes in five-dimensions}},
\newblock Nucl. Phys. B \textbf{531}, 239 (1998),
\newblock \doi{10.1016/S0550-3213(98)00604-X},
\newblock \eprint{hep-th/9805097}.

\bibitem{Ashtekar:1996cd}
A.~Ashtekar, J.~Bicak and B.~G. Schmidt,
\newblock \emph{{Asymptotic structure of symmetry reduced general relativity}},
\newblock Phys. Rev. D \textbf{55}, 669 (1997),
\newblock \doi{10.1103/PhysRevD.55.669},
\newblock \eprint{gr-qc/9608042}.

\bibitem{Barnich:2006av}
G.~Barnich and G.~Compere,
\newblock \emph{{Classical central extension for asymptotic symmetries at null infinity in three spacetime dimensions}},
\newblock Class. Quant. Grav. \textbf{24}, F15 (2007),
\newblock \doi{10.1088/0264-9381/24/5/F01},
\newblock \eprint{gr-qc/0610130}.

\bibitem{Compere:2017knf}
G.~Comp\`ere and A.~Fiorucci,
\newblock \emph{{Asymptotically flat spacetimes with BMS$_3$ symmetry}},
\newblock Class. Quant. Grav. \textbf{34}(20), 204002 (2017),
\newblock \doi{10.1088/1361-6382/aa8aad},
\newblock \eprint{1705.06217}.

\bibitem{Barnich:2012aw}
G.~Barnich, A.~Gomberoff and H.~A. Gonzalez,
\newblock \emph{{The Flat limit of three dimensional asymptotically anti-de Sitter spacetimes}},
\newblock Phys. Rev. D \textbf{86}, 024020 (2012),
\newblock \doi{10.1103/PhysRevD.86.024020},
\newblock \eprint{1204.3288}.

\bibitem{Cornalba:2002fi}
L.~Cornalba and M.~S. Costa,
\newblock \emph{{A New cosmological scenario in string theory}},
\newblock Phys. Rev. D \textbf{66}, 066001 (2002),
\newblock \doi{10.1103/PhysRevD.66.066001},
\newblock \eprint{hep-th/0203031}.

\bibitem{Cornalba:2003kd}
L.~Cornalba and M.~S. Costa,
\newblock \emph{{Time dependent orbifolds and string cosmology}},
\newblock Fortsch. Phys. \textbf{52}, 145 (2004),
\newblock \doi{10.1002/prop.200310123},
\newblock \eprint{hep-th/0310099}.

\bibitem{Barnich:2012xq}
G.~Barnich,
\newblock \emph{{Entropy of three-dimensional asymptotically flat cosmological solutions}},
\newblock JHEP \textbf{1210}, 095 (2012),
\newblock \doi{10.1007/JHEP10(2012)095},
\newblock \eprint{1208.4371}.

\bibitem{Bagchi:2012xr}
A.~Bagchi, S.~Detournay, R.~Fareghbal and J.~Simon,
\newblock \emph{{Holography of 3d Flat Cosmological Horizons}},
\newblock Phys.Rev.Lett. \textbf{110}, 141302 (2013),
\newblock \doi{10.1103/PhysRevLett.110.141302},
\newblock \eprint{1208.4372}.

\bibitem{Porfyriadis:2010vg}
A.~P. Porfyriadis and F.~Wilczek,
\newblock \emph{{Effective Action, Boundary Conditions, and Virasoro Algebra for AdS$_3$}}  (2010),
\newblock \eprint{1007.1031}.

\bibitem{Compere:2013aya}
G.~Compere, W.~Song and A.~Strominger,
\newblock \emph{{Chiral Liouville Gravity}},
\newblock JHEP \textbf{1305}, 154 (2013),
\newblock \doi{10.1007/JHEP05(2013)154},
\newblock \eprint{1303.2660}.

\bibitem{Afshar:2016wfy}
H.~Afshar, S.~Detournay, D.~Grumiller, W.~Merbis, A.~Perez, D.~Tempo and R.~Troncoso,
\newblock \emph{{Soft Heisenberg hair on black holes in three dimensions}},
\newblock Phys. Rev. D \textbf{93}(10), 101503 (2016),
\newblock \doi{10.1103/PhysRevD.93.101503},
\newblock \eprint{1603.04824}.

\bibitem{Troessaert:2013fma}
C.~Troessaert,
\newblock \emph{{Enhanced asymptotic symmetry algebra of $AdS_{3}$}},
\newblock JHEP \textbf{08}, 044 (2013),
\newblock \doi{10.1007/JHEP08(2013)044},
\newblock \eprint{1303.3296}.

\bibitem{Avery:2013dja}
S.~G. Avery, R.~R. Poojary and N.~V. Suryanarayana,
\newblock \emph{{An sl(2,$\mathbb{R}$) current algebra from $AdS_3$ gravity}},
\newblock JHEP \textbf{01}, 144 (2014),
\newblock \doi{10.1007/JHEP01(2014)144},
\newblock \eprint{1304.4252}.

\bibitem{Grumiller:2016pqb}
D.~Grumiller and M.~Riegler,
\newblock \emph{{Most general AdS$_{3}$ boundary conditions}},
\newblock JHEP \textbf{10}, 023 (2016),
\newblock \doi{10.1007/JHEP10(2016)023},
\newblock \eprint{1608.01308}.

\bibitem{Afshar:2016kjj}
H.~Afshar, D.~Grumiller, W.~Merbis, A.~Perez, D.~Tempo and R.~Troncoso,
\newblock \emph{{Soft hairy horizons in three spacetime dimensions}},
\newblock Phys. Rev. D \textbf{95}(10), 106005 (2017),
\newblock \doi{10.1103/PhysRevD.95.106005},
\newblock \eprint{1611.09783}.

\bibitem{Detournay:2016sfv}
S.~Detournay and M.~Riegler,
\newblock \emph{{Enhanced Asymptotic Symmetry Algebra of 2+1 Dimensional Flat Space}},
\newblock Phys. Rev. D \textbf{95}(4), 046008 (2017),
\newblock \doi{10.1103/PhysRevD.95.046008},
\newblock \eprint{1612.00278}.

\bibitem{Grumiller:2017sjh}
D.~Grumiller, W.~Merbis and M.~Riegler,
\newblock \emph{{Most general flat space boundary conditions in three-dimensional Einstein gravity}},
\newblock Class. Quant. Grav. \textbf{34}(18), 184001 (2017),
\newblock \doi{10.1088/1361-6382/aa8004},
\newblock \eprint{1704.07419}.

\bibitem{Grumiller:2019ygj}
D.~Grumiller, M.~M. Sheikh-Jabbari, C.~Troessaert and R.~Wutte,
\newblock \emph{{Interpolating Between Asymptotic and Near Horizon Symmetries}},
\newblock JHEP \textbf{03}, 035 (2020),
\newblock \doi{10.1007/JHEP03(2020)035},
\newblock \eprint{1911.04503}.

\bibitem{Fiola:1994ir}
T.~M. Fiola, J.~Preskill, A.~Strominger and S.~P. Trivedi,
\newblock \emph{{Black hole thermodynamics and information loss in two-dimensions}},
\newblock Phys. Rev. D \textbf{50}, 3987 (1994),
\newblock \doi{10.1103/PhysRevD.50.3987},
\newblock \eprint{hep-th/9403137}.

\bibitem{Maldacena:1998uz}
J.~M. Maldacena, J.~Michelson and A.~Strominger,
\newblock \emph{{Anti-de Sitter fragmentation}},
\newblock JHEP \textbf{02}, 011 (1999),
\newblock \doi{10.1088/1126-6708/1999/02/011},
\newblock \eprint{hep-th/9812073}.

\bibitem{Kunduri:2013gce}
H.~K. Kunduri and J.~Lucietti,
\newblock \emph{{Classification of near-horizon geometries of extremal black holes}},
\newblock Living Rev. Rel. \textbf{16}, 8 (2013),
\newblock \doi{10.12942/lrr-2013-8},
\newblock \eprint{1306.2517}.

\bibitem{Dunajski:2023xrd}
M.~Dunajski and J.~Lucietti,
\newblock \emph{{Intrinsic rigidity of extremal horizons}}  (2023),
\newblock \eprint{2306.17512}.

\bibitem{Anninos:2009yc}
D.~Anninos and T.~Hartman,
\newblock \emph{{Holography at an Extremal De Sitter Horizon}},
\newblock JHEP \textbf{03}, 096 (2010),
\newblock \doi{10.1007/JHEP03(2010)096},
\newblock \eprint{0910.4587}.

\bibitem{Almheiri:2014cka}
A.~Almheiri and J.~Polchinski,
\newblock \emph{{Models of AdS$_{2}$ backreaction and holography}},
\newblock JHEP \textbf{11}, 014 (2015),
\newblock \doi{10.1007/JHEP11(2015)014},
\newblock \eprint{1402.6334}.

\bibitem{Maldacena:2016upp}
J.~Maldacena, D.~Stanford and Z.~Yang,
\newblock \emph{{Conformal symmetry and its breaking in two dimensional Nearly Anti-de-Sitter space}},
\newblock PTEP \textbf{2016}(12), 12C104 (2016),
\newblock \doi{10.1093/ptep/ptw124},
\newblock \eprint{1606.01857}.

\bibitem{Sarosi:2017ykf}
G.~S\'arosi,
\newblock \emph{{AdS$_{2}$ holography and the SYK model}},
\newblock PoS \textbf{Modave2017}, 001 (2018),
\newblock \doi{10.22323/1.323.0001},
\newblock \eprint{1711.08482}.

\bibitem{Mertens:2022irh}
T.~G. Mertens and G.~J. Turiaci,
\newblock \emph{{Solvable models of quantum black holes: a review on Jackiw\textendash{}Teitelboim gravity}},
\newblock Living Rev. Rel. \textbf{26}(1), 4 (2023),
\newblock \doi{10.1007/s41114-023-00046-1},
\newblock \eprint{2210.10846}.

\bibitem{Castro:2021csm}
A.~Castro, V.~Godet, J.~Sim\'on, W.~Song and B.~Yu,
\newblock \emph{{Gravitational perturbations from NHEK to Kerr}},
\newblock JHEP \textbf{07}, 218 (2021),
\newblock \doi{10.1007/JHEP07(2021)218},
\newblock \eprint{2102.08060}.

\bibitem{Teitelboim:1983ux}
C.~Teitelboim,
\newblock \emph{{Gravitation and Hamiltonian Structure in Two Space-Time Dimensions}},
\newblock Phys. Lett. \textbf{126B}, 41 (1983),
\newblock \doi{10.1016/0370-2693(83)90012-6}.

\bibitem{Jackiw:1984je}
R.~Jackiw,
\newblock \emph{{Lower Dimensional Gravity}},
\newblock Nucl. Phys. \textbf{B252}, 343 (1985),
\newblock \doi{10.1016/0550-3213(85)90448-1}.

\bibitem{Hotta:1998iq}
M.~Hotta,
\newblock \emph{{Asymptotic isometry and two-dimensional anti-de Sitter gravity}}  (1998),
\newblock \eprint{gr-qc/9809035}.

\bibitem{Cadoni:1999ja}
M.~Cadoni and S.~Mignemi,
\newblock \emph{{Asymptotic symmetries of AdS(2) and conformal group in d = 1}},
\newblock Nucl. Phys. B \textbf{557}, 165 (1999),
\newblock \doi{10.1016/S0550-3213(99)00398-3},
\newblock \eprint{hep-th/9902040}.

\bibitem{Navarro-Salas:1999zer}
J.~Navarro-Salas and P.~Navarro,
\newblock \emph{{AdS(2) / CFT(1) correspondence and near extremal black hole entropy}},
\newblock Nucl. Phys. B \textbf{579}, 250 (2000),
\newblock \doi{10.1016/S0550-3213(00)00165-6},
\newblock \eprint{hep-th/9910076}.

\bibitem{Grumiller:2017qao}
D.~Grumiller, R.~McNees, J.~Salzer, C.~Valc\'arcel and D.~Vassilevich,
\newblock \emph{{Menagerie of AdS$_{2}$ boundary conditions}},
\newblock JHEP \textbf{10}, 203 (2017),
\newblock \doi{10.1007/JHEP10(2017)203},
\newblock \eprint{1708.08471}.

\bibitem{Godet:2020xpk}
V.~Godet and C.~Marteau,
\newblock \emph{{New boundary conditions for AdS$_{2}$}},
\newblock JHEP \textbf{12}, 020 (2020),
\newblock \doi{10.1007/JHEP12(2020)020},
\newblock \eprint{2005.08999}.

\bibitem{Hofman:2011zj}
D.~M. Hofman and A.~Strominger,
\newblock \emph{{Chiral Scale and Conformal Invariance in 2D Quantum Field Theory}},
\newblock Phys.Rev.Lett. \textbf{107}, 161601 (2011),
\newblock \doi{10.1103/PhysRevLett.107.161601},
\newblock \eprint{1107.2917}.

\bibitem{Detournay:2012pc}
S.~Detournay, T.~Hartman and D.~M. Hofman,
\newblock \emph{{Warped Conformal Field Theory}},
\newblock Phys.Rev. \textbf{D86}, 124018 (2012),
\newblock \doi{10.1103/PhysRevD.86.124018},
\newblock \eprint{1210.0539}.

\bibitem{Castro:2015uaa}
A.~Castro, D.~M. Hofman and G.~S\'arosi,
\newblock \emph{{Warped Weyl fermion partition functions}},
\newblock JHEP \textbf{11}, 129 (2015),
\newblock \doi{10.1007/JHEP11(2015)129},
\newblock \eprint{1508.06302}.

\bibitem{Castro:2015csg}
A.~Castro, D.~M. Hofman and N.~Iqbal,
\newblock \emph{{Entanglement Entropy in Warped Conformal Field Theories}},
\newblock JHEP \textbf{02}, 033 (2016),
\newblock \doi{10.1007/JHEP02(2016)033},
\newblock \eprint{1511.00707}.

\bibitem{Song:2017czq}
W.~Song and J.~Xu,
\newblock \emph{{Correlation Functions of Warped CFT}},
\newblock JHEP \textbf{04}, 067 (2018),
\newblock \doi{10.1007/JHEP04(2018)067},
\newblock \eprint{1706.07621}.

\bibitem{Apolo:2018eky}
L.~Apolo and W.~Song,
\newblock \emph{{Bootstrapping holographic warped CFTs or: how I learned to stop worrying and tolerate negative norms}},
\newblock JHEP \textbf{07}, 112 (2018),
\newblock \doi{10.1007/JHEP07(2018)112},
\newblock \eprint{1804.10525}.

\bibitem{Chaturvedi:2018uov}
P.~Chaturvedi, Y.~Gu, W.~Song and B.~Yu,
\newblock \emph{{A note on the complex SYK model and warped CFTs}},
\newblock JHEP \textbf{12}, 101 (2018),
\newblock \doi{10.1007/JHEP12(2018)101},
\newblock \eprint{1808.08062}.

\bibitem{Aggarwal:2022xfd}
A.~Aggarwal, A.~Castro, S.~Detournay and B.~M\"uhlmann,
\newblock \emph{{Near-extremal limits of warped CFTs}},
\newblock SciPost Phys. \textbf{15}(2), 056 (2023),
\newblock \doi{10.21468/SciPostPhys.15.2.056},
\newblock \eprint{2211.03770}.

\bibitem{Afshar:2019axx}
H.~Afshar, H.~A. Gonz\'alez, D.~Grumiller and D.~Vassilevich,
\newblock \emph{{Flat space holography and the complex Sachdev-Ye-Kitaev model}},
\newblock Phys. Rev. D \textbf{101}(8), 086024 (2020),
\newblock \doi{10.1103/PhysRevD.101.086024},
\newblock \eprint{1911.05739}.

\bibitem{Guica:2008mu}
M.~Guica, T.~Hartman, W.~Song and A.~Strominger,
\newblock \emph{{The Kerr/CFT Correspondence}},
\newblock Phys. Rev. D \textbf{80}, 124008 (2009),
\newblock \doi{10.1103/PhysRevD.80.124008},
\newblock \eprint{0809.4266}.

\bibitem{Bardeen:1999px}
J.~M. Bardeen and G.~T. Horowitz,
\newblock \emph{{The Extreme Kerr throat geometry: A Vacuum analog of AdS(2) x S**2}},
\newblock Phys. Rev. D \textbf{60}, 104030 (1999),
\newblock \doi{10.1103/PhysRevD.60.104030},
\newblock \eprint{hep-th/9905099}.

\bibitem{Moussa:2003fc}
K.~A. Moussa, G.~Clement and C.~Leygnac,
\newblock \emph{{The Black holes of topologically massive gravity}},
\newblock Class.Quant.Grav. \textbf{20}, L277 (2003),
\newblock \doi{10.1088/0264-9381/20/24/L01},
\newblock \eprint{gr-qc/0303042}.

\bibitem{Bouchareb:2007yx}
A.~Bouchareb and G.~Clement,
\newblock \emph{{Black hole mass and angular momentum in topologically massive gravity}},
\newblock Class.Quant.Grav. \textbf{24}, 5581 (2007),
\newblock \doi{10.1088/0264-9381/24/22/018},
\newblock \eprint{0706.0263}.

\bibitem{Banados:2005da}
M.~Banados, G.~Barnich, G.~Compere and A.~Gomberoff,
\newblock \emph{{Three dimensional origin of Godel spacetimes and black holes}},
\newblock Phys. Rev. D \textbf{73}, 044006 (2006),
\newblock \doi{10.1103/PhysRevD.73.044006},
\newblock \eprint{hep-th/0512105}.

\bibitem{Detournay:2012dz}
S.~Detournay and M.~Guica,
\newblock \emph{{Stringy Schr\"odinger truncations}},
\newblock JHEP \textbf{08}, 121 (2013),
\newblock \doi{10.1007/JHEP08(2013)121},
\newblock \eprint{1212.6792}.

\bibitem{Tonni:2010gb}
E.~Tonni,
\newblock \emph{{Warped black holes in 3D general massive gravity}},
\newblock JHEP \textbf{08}, 070 (2010),
\newblock \doi{10.1007/JHEP08(2010)070},
\newblock \eprint{1006.3489}.

\bibitem{Donnay:2015iia}
L.~Donnay and G.~Giribet,
\newblock \emph{{Holographic entropy of Warped-AdS$_{3}$ black holes}},
\newblock JHEP \textbf{06}, 099 (2015),
\newblock \doi{10.1007/JHEP06(2015)099},
\newblock \eprint{1504.05640}.

\bibitem{Detournay:2016gao}
S.~Detournay, L.-A. Douxchamps, G.~S. Ng and C.~Zwikel,
\newblock \emph{{Warped AdS$_{3}$ black holes in higher derivative gravity theories}},
\newblock JHEP \textbf{06}, 014 (2016),
\newblock \doi{10.1007/JHEP06(2016)014},
\newblock \eprint{1602.09089}.

\bibitem{Compere:2007in}
G.~Compere and S.~Detournay,
\newblock \emph{{Centrally extended symmetry algebra of asymptotically Godel spacetimes}},
\newblock JHEP \textbf{03}, 098 (2007),
\newblock \doi{10.1088/1126-6708/2007/03/098},
\newblock \eprint{hep-th/0701039}.

\bibitem{Compere:2008cv}
G.~Compere and S.~Detournay,
\newblock \emph{{Semi-classical central charge in topologically massive gravity}},
\newblock Class.Quant.Grav. \textbf{26}, 012001 (2009),
\newblock \doi{10.1088/0264-9381/26/1/012001, 10.1088/0264-9381/26/13/139801},
\newblock \eprint{0808.1911}.

\bibitem{Compere:2009zj}
G.~Compere and S.~Detournay,
\newblock \emph{{Boundary conditions for spacelike and timelike warped $AdS_{3}$ spaces in topologically massive gravity}},
\newblock JHEP \textbf{08}, 092 (2009),
\newblock \doi{10.1088/1126-6708/2009/08/092},
\newblock \eprint{0906.1243}.

\bibitem{Henneaux:2011hv}
M.~Henneaux, C.~Martinez and R.~Troncoso,
\newblock \emph{{Asymptotically warped anti-de Sitter spacetimes in topologically massive gravity}},
\newblock Phys. Rev. D \textbf{84}, 124016 (2011),
\newblock \doi{10.1103/PhysRevD.84.124016},
\newblock \eprint{1108.2841}.

\bibitem{Song:2016gtd}
W.~Song, Q.~Wen and J.~Xu,
\newblock \emph{{Modifications to Holographic Entanglement Entropy in Warped CFT}},
\newblock JHEP \textbf{02}, 067 (2017),
\newblock \doi{10.1007/JHEP02(2017)067},
\newblock \eprint{1610.00727}.

\bibitem{Azeyanagi:2018har}
T.~Azeyanagi, S.~Detournay and M.~Riegler,
\newblock \emph{{Warped Black Holes in Lower-Spin Gravity}},
\newblock Phys. Rev. D \textbf{99}(2), 026013 (2019),
\newblock \doi{10.1103/PhysRevD.99.026013},
\newblock \eprint{1801.07263}.

\bibitem{Aggarwal:2020igb}
A.~Aggarwal, L.~Ciambelli, S.~Detournay and A.~Somerhausen,
\newblock \emph{{Boundary conditions for warped AdS$_{3}$ in quadratic ensemble}},
\newblock JHEP \textbf{22}, 013 (2020),
\newblock \doi{10.1007/JHEP05(2022)013},
\newblock \eprint{2112.13116}.

\bibitem{Anninos:2008fx}
D.~Anninos, W.~Li, M.~Padi, W.~Song and A.~Strominger,
\newblock \emph{{Warped AdS(3) Black Holes}},
\newblock JHEP \textbf{0903}, 130 (2009),
\newblock \doi{10.1088/1126-6708/2009/03/130},
\newblock \eprint{0807.3040}.

\bibitem{Chen:2009cg}
B.~Chen, B.~Ning and Z.-b. Xu,
\newblock \emph{{Real-time correlators in warped AdS/CFT correspondence}},
\newblock JHEP \textbf{02}, 031 (2010),
\newblock \doi{10.1007/JHEP02(2010)031},
\newblock \eprint{0911.0167}.

\bibitem{Song:2016pwx}
W.~Song, Q.~Wen and J.~Xu,
\newblock \emph{{Generalized Gravitational Entropy for Warped Anti\textendash{}de Sitter Space}},
\newblock Phys. Rev. Lett. \textbf{117}(1), 011602 (2016),
\newblock \doi{10.1103/PhysRevLett.117.011602},
\newblock \eprint{1601.02634}.

\bibitem{Compere:2014bia}
G.~Comp\`ere, M.~Guica and M.~J. Rodriguez,
\newblock \emph{{Two Virasoro symmetries in stringy warped AdS$_{3}$}},
\newblock JHEP \textbf{12}, 012 (2014),
\newblock \doi{10.1007/JHEP12(2014)012},
\newblock \eprint{1407.7871}.

\bibitem{Guica:2013jza}
M.~Guica,
\newblock \emph{{Decrypting the warped black strings}},
\newblock JHEP \textbf{11}, 025 (2013),
\newblock \doi{10.1007/JHEP11(2013)025},
\newblock \eprint{1305.7249}.

\bibitem{El-Showk:2011euy}
S.~El-Showk and M.~Guica,
\newblock \emph{{Kerr/CFT, dipole theories and nonrelativistic CFTs}},
\newblock JHEP \textbf{12}, 009 (2012),
\newblock \doi{10.1007/JHEP12(2012)009},
\newblock \eprint{1108.6091}.

\bibitem{Song:2011sr}
W.~Song and A.~Strominger,
\newblock \emph{{Warped AdS3/Dipole-CFT Duality}},
\newblock JHEP \textbf{1205}, 120 (2012),
\newblock \doi{10.1007/JHEP05(2012)120},
\newblock \eprint{1109.0544}.

\bibitem{Guica:2017lia}
M.~Guica,
\newblock \emph{{An integrable Lorentz-breaking deformation of two-dimensional CFTs}},
\newblock SciPost Phys. \textbf{5}(5), 048 (2018),
\newblock \doi{10.21468/SciPostPhys.5.5.048},
\newblock \eprint{1710.08415}.

\bibitem{Bzowski:2018pcy}
A.~Bzowski and M.~Guica,
\newblock \emph{{The holographic interpretation of $J \bar T$-deformed CFTs}},
\newblock JHEP \textbf{01}, 198 (2019),
\newblock \doi{10.1007/JHEP01(2019)198},
\newblock \eprint{1803.09753}.

\bibitem{Aggarwal:2023peg}
A.~Aggarwal, A.~Castro, S.~Detournay and B.~M\"uhlmann,
\newblock \emph{{Near-Extremal Limits of Warped Black Holes}},
\newblock SciPost Phys. \textbf{15}, 083 (2023),
\newblock \doi{10.21468/SciPostPhys.15.3.083},
\newblock \eprint{2304.10102}.

\bibitem{Aggarwal:2019iay}
A.~Aggarwal, A.~Castro and S.~Detournay,
\newblock \emph{{Warped Symmetries of the Kerr Black Hole}},
\newblock JHEP \textbf{01}, 016 (2020),
\newblock \doi{10.1007/JHEP01(2020)016},
\newblock \eprint{1909.03137}.

\bibitem{Haco:2018ske}
S.~Haco, S.~W. Hawking, M.~J. Perry and A.~Strominger,
\newblock \emph{{Black Hole Entropy and Soft Hair}},
\newblock JHEP \textbf{12}, 098 (2018),
\newblock \doi{10.1007/JHEP12(2018)098},
\newblock \eprint{1810.01847}.

\bibitem{Rakic:2023vhv}
I.~Rakic, M.~Rangamani and G.~J. Turiaci,
\newblock \emph{{Thermodynamics of the near-extremal Kerr spacetime}}  (2023),
\newblock \eprint{2310.04532}.

\bibitem{Kapec:2023ruw}
D.~Kapec, A.~Sheta, A.~Strominger and C.~Toldo,
\newblock \emph{{Logarithmic Corrections to Kerr Thermodynamics}}  (2023),
\newblock \eprint{2310.00848}.

\bibitem{Matsuo:2009sj}
Y.~Matsuo, T.~Tsukioka and C.-M. Yoo,
\newblock \emph{{Another Realization of Kerr/CFT Correspondence}},
\newblock Nucl. Phys. B \textbf{825}, 231 (2010),
\newblock \doi{10.1016/j.nuclphysb.2009.09.025},
\newblock \eprint{0907.0303}.

\bibitem{Matsuo:2009pg}
Y.~Matsuo, T.~Tsukioka and C.-M. Yoo,
\newblock \emph{{Yet Another Realization of Kerr/CFT Correspondence}},
\newblock EPL \textbf{89}(6), 60001 (2010),
\newblock \doi{10.1209/0295-5075/89/60001},
\newblock \eprint{0907.4272}.

\bibitem{Cadoni:1998sg}
M.~Cadoni and S.~Mignemi,
\newblock \emph{{Entropy of 2-D black holes from counting microstates}},
\newblock Phys. Rev. D \textbf{59}, 081501 (1999),
\newblock \doi{10.1103/PhysRevD.59.081501},
\newblock \eprint{hep-th/9810251}.

\bibitem{Afshar:2015wjm}
H.~Afshar, S.~Detournay, D.~Grumiller and B.~Oblak,
\newblock \emph{{Near-Horizon Geometry and Warped Conformal Symmetry}},
\newblock JHEP \textbf{03}, 187 (2016),
\newblock \doi{10.1007/JHEP03(2016)187},
\newblock \eprint{1512.08233}.

\bibitem{Balasubramanian:2009bg}
V.~Balasubramanian, J.~de~Boer, M.~M. Sheikh-Jabbari and J.~Simon,
\newblock \emph{{What is a chiral 2d CFT? And what does it have to do with extremal black holes?}},
\newblock JHEP \textbf{02}, 017 (2010),
\newblock \doi{10.1007/JHEP02(2010)017},
\newblock \eprint{0906.3272}.

\bibitem{Castro:2022cuo}
A.~Castro, F.~Mariani and C.~Toldo,
\newblock \emph{{Near-extremal limits of de Sitter black holes}},
\newblock JHEP \textbf{07}, 131 (2023),
\newblock \doi{10.1007/JHEP07(2023)131},
\newblock \eprint{2212.14356}.

\bibitem{Afshar:2021qvi}
H.~Afshar and B.~Oblak,
\newblock \emph{{Flat JT gravity and the BMS-Schwarzian}},
\newblock JHEP \textbf{11}, 172 (2022),
\newblock \doi{10.1007/JHEP11(2022)172},
\newblock \eprint{2112.14609}.

\bibitem{deBoer:2010ac}
J.~de~Boer, M.~M. Sheikh-Jabbari and J.~Simon,
\newblock \emph{{Near Horizon Limits of Massless BTZ and Their CFT Duals}},
\newblock Class. Quant. Grav. \textbf{28}, 175012 (2011),
\newblock \doi{10.1088/0264-9381/28/17/175012},
\newblock \eprint{1011.1897}.

\bibitem{Compere:2019qed}
G.~Comp\`ere,
\newblock \emph{{Advanced Lectures on General Relativity}}, vol. 952,
\newblock Springer, Cham, Cham, Switzerland,
\newblock ISBN 978-3-030-04259-2, 978-3-030-04260-8,
\newblock \doi{10.1007/978-3-030-04260-8} (2019).

\bibitem{Kapec:2019hro}
D.~Kapec and A.~Lupsasca,
\newblock \emph{{Particle motion near high-spin black holes}},
\newblock Class. Quant. Grav. \textbf{37}(1), 015006 (2020),
\newblock \doi{10.1088/1361-6382/ab519e},
\newblock \eprint{1905.11406}.

\bibitem{Castro:2019crn}
A.~Castro and V.~Godet,
\newblock \emph{{Breaking away from the near horizon of extreme Kerr}},
\newblock SciPost Phys. \textbf{8}(6), 089 (2020),
\newblock \doi{10.21468/SciPostPhys.8.6.089},
\newblock \eprint{1906.09083}.

\bibitem{Ghosh:2019rcj}
A.~Ghosh, H.~Maxfield and G.~J. Turiaci,
\newblock \emph{{A universal Schwarzian sector in two-dimensional conformal field theories}},
\newblock JHEP \textbf{05}, 104 (2020),
\newblock \doi{10.1007/JHEP05(2020)104},
\newblock \eprint{1912.07654}.

\bibitem{Iliesiu:2020qvm}
L.~V. Iliesiu and G.~J. Turiaci,
\newblock \emph{{The statistical mechanics of near-extremal black holes}},
\newblock JHEP \textbf{05}, 145 (2021),
\newblock \doi{10.1007/JHEP05(2021)145},
\newblock \eprint{2003.02860}.

\end{thebibliography}

%%%%%%%%%% END TODO: BIBLIOGRAPHY

\end{document}